\documentclass[conference]{IEEEtran}

\usepackage[T1]{fontenc}
\usepackage[table]{xcolor}
\usepackage{hyperref}
\usepackage{amsmath}
\usepackage{graphicx}
\usepackage{amssymb}
\usepackage{amsthm}
\usepackage{amsmath}
\usepackage{booktabs}
\usepackage{pifont}
\usepackage{algpseudocode}
\usepackage{tabularx}
\usepackage[available,functional,reproduced]{ndssbadges}
\usepackage{cite}
\usepackage{enumitem}
\let\OLDthebibliography\thebibliography
\renewcommand\thebibliography[1]{%
  \OLDthebibliography{#1}%
  \setlength{\itemsep}{0pt plus 0.2ex}%
  \setlength{\parskip}{0pt}%
}
\newcommand{\cmark}{\ding{51}}
\newcommand{\xmark}{\ding{55}}

\newtheorem{theorem}{Theorem}
\newtheorem{corollary}{Corollary}

\begin{document}

\title{zk-DEAP: A Zero-Knowledge Decentralized Encrypted Aggregation Protocol}

\author{
    \IEEEauthorblockN{
        Jude Ramanan,
        Georgiana Haldeman,
        and Noah Apthorpe
    }
    \IEEEauthorblockA{
        Colgate University\\
        \{jramanan, ghaldeman, napthorpe\}@colgate.edu
    }
}

\IEEEoverridecommandlockouts
\makeatletter\def\@IEEEpubidpullup{6.5\baselineskip}\makeatother
\IEEEpubid{\parbox{\columnwidth}{
		Network and Distributed System Security (NDSS) Symposium 2027\\
		22 - 26 March 2027, Seoul, Republic of Korea\\
		ISBN 978-1-970672-09-1\\  
		https://dx.doi.org/10.14722/ndss.2027.230198\\
		www.ndss-symposium.org
}
\hspace{\columnsep}\makebox[\columnwidth]{}}

\maketitle

\begin{abstract}
Secure aggregation in domains such as federated learning, Internet of Things, and voting requires operating on private inputs while preserving confidentiality, verifying that each input satisfies constraints, and avoiding trusted parties. Existing protocols achieve at most two of these requirements. We present zk-DEAP, a protocol combining lifted ElGamal homomorphic encryption, FROST threshold cryptography, and zero-knowledge proofs (Bulletproofs, zk-SNARKs, or zk-STARKs) for fully decentralized, verifiable, privacy-preserving aggregation.
Our core technical contribution is a cryptographic binding mechanism preventing substitution attacks, where a valid proof for one input is paired with a ciphertext encrypting a different, malicious value. For Bulletproofs, we bind via ciphertext-derived randomness modification. 
For zk-SNARKs/zk-STARKs, to bind across algebraic structures (a well-known challenge in practical cryptography), we introduce a novel blinded arithmetic hash commitment with an embedded Schnorr constraint. These binding techniques force the ZKP's private witness to equal the encrypted value, ensuring any circuit-verified constraint applies to the ciphertext.
We implement three production-ready, defense-in-depth variants in Rust and evaluate across 5--500 participants. Compared to unverified aggregation, zk-DEAP adds modest overhead while providing cryptographic guarantees against input manipulation that statistical defenses cannot match. 
\end{abstract}

\IEEEpeerreviewmaketitle

\section{Introduction} \label{section:intro}

Secure aggregation enables multiple participants to combine their private inputs into an aggregate result without revealing individual contributions. This is critical in federated learning (FL), where participants train models locally and submit gradient updates for aggregation into a global model~\cite{mcmahan2017communication}. Three security requirements arise simultaneously. First, individual updates must remain confidential, as gradient updates can leak sensitive training data through inversion attacks~\cite{zhu2019deep}. Second, each update must satisfy constraints to prevent model poisoning, as a single unconstrained gradient can corrupt the global model~\cite{blanchard2017machine, bagdasaryan2020backdoor}. Third, no single aggregator should be trusted with all updates, as centralized aggregators create a single point of failure for both confidentiality and integrity. Parallel requirements arise in smart grid telemetry~\cite{ZHANG2022103118},  voting~\cite{cramer1997}, blockchains~\cite{10292604}, and Internet of Things (IoT) sensor aggregation~\cite{privda2021}, making secure decentralized aggregation a widespread problem in zero-trust and least-privileged cryptography.
    
Combining off-the-shelf homomorphic encryption (HE) with zero-knowledge proofs (ZKPs) appears to solve the first two requirements. HE enables aggregation over encrypted inputs, while ZKPs verify constraints without decryption. However, this combination is vulnerable to substitution attacks. In a \textit{value substitution} attack, a malicious participant generates a valid ZKP proving that its input $s_i$ satisfies the constraint predicate $\Phi$ (for example, bounded gradient norm or cosine-similarity check), but pairs it with a ciphertext encrypting a completely different $s'_i$ that violates $\Phi$. Because the ZKP and ciphertext are independent cryptographic objects, verifiers cannot detect the mismatch. The malicious ciphertext is aggregated with honest contributions, silently corrupting the result. For instance, if $\Phi$ requires $s_i \in [0,255]$, an adversary can encrypt $10^5$ (which violates $\Phi$) while presenting a valid range proof for $s_i=34$ (which satisfies $\Phi$), thereby corrupting the aggregate. A subtler variant, \textit{randomness substitution}, targets ciphertext consistency. In multi-component encryption schemes, a malicious participant can submit a valid ZKP bound to one ciphertext component but pair it with a malformed companion component, causing threshold decryption to fail or produce an incorrect aggregate.

\begin{figure}[t]
\centering
\includegraphics[width=0.25\textwidth]{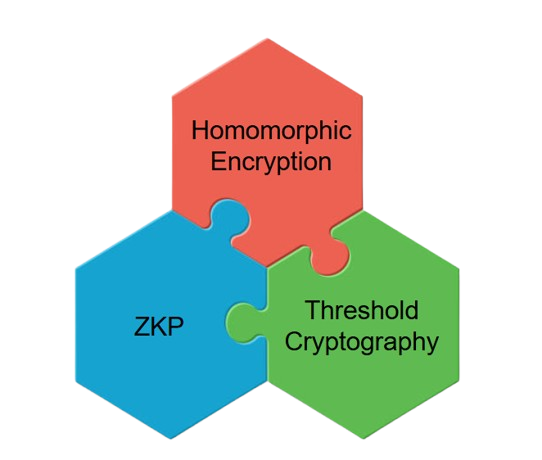}
    \caption{Decentralized Secure Zero-Trust Aggregation Puzzle: (1) Homomorphic aggregation prevents individual inputs from being revealed. (2) Zero-knowledge proofs validate constraints on the inputs to ensure they are not malicious. (3) Threshold cryptography allows the inputs and aggregate to be safely encrypted and decrypted.} 
    \label{fig:puzzle}
\end{figure}

These substitution attacks can occur anywhere the proof and ciphertext are different mathematical objects. They do not arise in commitment-based single-server aggregation (RoFL~\cite{rofl2023}, ACORN~\cite{acorn2023}, EIFFeL~\cite{eiffel2022}). In these schemes, the ZKP operates directly over a Pedersen commitment~\cite{pedersen1991non} in the same algebraic group, and there is no separate ciphertext to substitute against. This is not an option in the decentralized setting, however, as threshold ElGamal decryption~\cite{elgamal1985, cramer1997}, with keys established through Flexible Round-Optimized Schnorr Threshold (FROST) distributed key generation~\cite{KomloG20}, requires a two-component ciphertext distinct from the ZKP witness domain. But flexible ZKP systems, like zero-knowledge Succinct Non-interactive ARgument of Knowledge (zk-SNARK)~\cite{bitansky2012extractable} and zero-knowledge Scalable Transparent ARgument of Knowledge (zk-STARK)~\cite{ben2018scalable}, operate over structures incompatible with the scalar field of Ristretto255~\cite{rfc9496}, the elliptic curve group used for FROST and ElGamal. This is a manifestation of the \textit{cross-domain binding problem}. Ensuring that a constraint verified in the ZKP’s field applies to the value encrypted in the ciphertext’s group without leaking the value or requiring expensive non-native field arithmetic is an open problem in cross-domain proof composition. The substitution attack surface arises in systems simultaneously requiring privacy, decentralization, and arbitrary-constraint verifiability, and is the target of our theoretical binding contributions.

Regarding decentralization, threshold cryptography and distributed key generation (DKG) allow HE to operate without trust relations by establishing shared encryption keys where any $t$ participants can collectively decrypt~\cite{GennaroJareckiKrawczykRabin07}. The ideal secure aggregation scheme in a distributed system can therefore be viewed as a 3-piece puzzle involving ZKPs, HE, and threshold cryptography (Figure~\ref{fig:puzzle}).
Recent work in secure aggregation~\cite{privda2021,baharon2024secure,deoliveira2024dcnets} has successfully integrated two pieces of this puzzle, but none has combined all three. 

Existing approaches that link ZKPs and HE, such as commit-and-prove SNARKs and Bulletproof-linked ElGamal variants~\cite{campanelli2019legosnark,lee2019saver,bunz2020zether}, assume the commitment and the proof are in the same field or pairing. This does not hold in every context, such as when FROST ties the ciphertext to Ristretto255 but the proofs use an unrelated prime field. 

We therefore propose and implement zk-DEAP, a Zero-Knowledge Decentralized Encrypted Aggregation Protocol.\footnote{\href{https://doi.org/10.6084/m9.figshare.33411562} {https://doi.org/10.6084/m9.figshare.33411562}} zk-DEAP is the first general-purpose secure aggregation scheme achieving simultaneous privacy, verifiability, decentralization, and anti-substitution binding for practical aggregation, with three different variants. zk-DEAP includes a novel binding mechanism that addresses the cross-domain binding problem without the need for foreign-field arithmetic in the circuit. Table~\ref{tab:related} compares zk-DEAP against existing secure aggregation protocols.

\begin{table*}[!tbp]
\centering
\caption{Comparison with existing secure aggregation protocols.
Priv.\ = Indistinguishability under chosen-plaintext attack input privacy;
Verif.\ = cryptographic input constraint verification;
Decen.\ = no single party whose absence prevents completion.
Values $d$, $b$, and $k$ are input dimension, bit length of input bound, and number of projections, respectively.}
\label{tab:related}
\begin{tabular}{llcccllll}
\toprule
\textbf{Protocol} & \textbf{Year} & \textbf{Priv.} & \textbf{Verif.} & \textbf{Decen.} & \textbf{Crypto Primitives} & \textbf{Trust Model} & \textbf{Proof Size} & \textbf{Context} \\
\midrule
SecAgg~\cite{bonawitz2017secagg} & 2017 & \cmark & \xmark & \xmark & Secret sharing, DH & Single server & --- & FL \\
SecAgg+~\cite{bell2022secaggplus} & 2020 & \cmark & \xmark & \xmark & Sparse secret sharing & Single server & --- & FL \\
PrivDA~\cite{privda2021} & 2021 & \cmark & \xmark & \xmark & Paillier HE, Ethereum & Trusted aggr. & --- & IoT \\
VeriFL~\cite{verifl2021} & 2021 & \cmark & \xmark$^\dag$ & \xmark & Homo.\ hash, secret sharing & Single server & --- & FL \\
EIFFeL~\cite{eiffel2022} & 2022 & \cmark & \cmark & \xmark & Secret sharing, SNIP & Server + clients & $^\ddag$ & FL \\
Zhang et al.~\cite{ZHANG2022103118} & 2022 & \cmark & \cmark & \xmark & HE + ZKP + proxy re-enc. & Trusted aggr. & Not Reported & Smart grid \\
Flamingo~\cite{flamingo2023} & 2023 & \cmark & \xmark & \xmark & DKG, ElGamal masking & Single server & --- & FL \\
RoFL~\cite{rofl2023} & 2023 & \cmark & \cmark & \xmark & Commitments, Bulletproofs & Single server & $O(d\log b)$ & FL \\
ELSA~\cite{elsa2023} & 2023 & \cmark & \cmark & \xmark$^\S$ & MPC, distributed trust & Two servers & $^\ddag$ & FL \\
ACORN~\cite{acorn2023} & 2023 & \cmark & \cmark & \xmark & Pedersen vector comm., IP arg. & Single server & $O(\log d)$ & FL \\
PPVDA~\cite{wang2024ppvda} & 2024 & \cmark & \xmark & \cmark & Homo.\ MAC, secret sharing & Peer-to-peer & --- & IoV \\
TrustDFL~\cite{wang2024trustdfl} & 2024 & \xmark & \cmark & \xmark & Groth16 ZKP, blockchain & Blockchain & 1,200\,B & FL \\
RiseFL~\cite{risefl2024} & 2024 & \cmark & \cmark & \xmark & Hybrid comm., random proj. & Single server & $O(k)$ & FL \\
GVSA~\cite{zhou2025gvsa} & 2025 & \xmark & \xmark & \cmark & Secret sharing, verif.\ tags & Peer-to-peer & --- & FL \\
LZKSA~\cite{lzksa2025} & 2025 & \cmark & \cmark & \xmark & Lattice ZKPs, RLWE & Single server & $O(d)$ & FL \\
\midrule
\textbf{zk-DEAP (BP)} & 2026 & \cmark & \cmark & \cmark & ElGamal + Bulletproofs + FROST & Peer-to-peer & 924\,B & Generic \\
\textbf{zk-DEAP (SNARK)} & 2026 & \cmark & \cmark & \cmark & ElGamal + Halo2 + FROST & Peer-to-peer & 1,981\,B & Generic \\
\textbf{zk-DEAP (STARK)} & 2026 & \cmark & \cmark & \cmark & ElGamal + Winterfell + FROST & Peer-to-peer & 18,803\,B & Generic \\
\bottomrule
\multicolumn{9}{l}{\footnotesize $^\dag$VeriFL verifies server aggregation integrity (correct summation), not client input constraints.} \\
\multicolumn{9}{l}{\footnotesize $^\ddag$EIFFeL/ELSA use information-theoretic proofs on secret-shared data; proof size not directly comparable.} \\
\multicolumn{9}{l}{\footnotesize $^\S$ELSA distributes trust across two servers but requires both to be available.} \\
\end{tabular}
\end{table*}

At the theoretical level, zk-DEAP uses a blinded arithmetic hash with embedded Schnorr constraints to address the cross-domain binding problem through a pre-challenge commitment that creates a cryptographic dependency between the ZKP's private witness and the Schnorr proof's extracted value. Modular reduction between the curve's scalar field and the proof system's field is handled uniformly in both the SNARK and STARK variants by an explicit carry value whose bit width tracks the input domain.

At the protocol implementation level, zk-DEAP first involves a setup phase to establish the public and private keys required for the protocol using FROST~\cite{KomloG20}. Next, all participants construct proof packages, each comprising an input encrypted under lifted ElGamal~\cite{cramer1997}, a binding Schnorr proof~\cite{schnorr1991efficient}, a zero-knowledge proof (Bulletproof~\cite{DBLP:conf/sp/BunzBBPCP18}, zk-SNARK~\cite{bitansky2012extractable}, or zk-STARK~\cite{ben2018scalable}), and an Ed25519 signature~\cite{bernstein2012highs, fips186-5}. These packages are then broadcast to the other participants, verified, and aggregated. Once it is time to decrypt the final aggregate, each participant performs a partial decryption, which is combined with Lagrange interpolation. Additional defense-in-depth and production features are also included to ensure robustness and real-world applicability.

At the evaluation level, analysis on networks of 5--500 participants demonstrates our protocol adds modest overhead to unverified aggregation. The Bulletproof variant adds 3.4\,ms of per-peer cryptographic overhead (2.2\,ms prove $+$ 1.2\,ms verify), with 924-byte proof packages. For $n = 100$ participants, total overhead per round remains approximately 153.0\,ms, representing approximately 0.25\% of a typical 60-second federated learning round~\cite{bonawitz2019towards}. The one-time FROST DKG cost, while exhibiting $O(n^2 t + n t^2)$ growth, is amortized across protocol deployments lasting hours or days. Communication overhead is $O(n^2)$, with per-participant bandwidth ranging from 92\,KB (Bulletproofs) to 1.9\,MB (zk-STARKs) for $n = 100$.

In summary, our contributions are:

(1) A novel cryptographic binding mechanism that prevents substitution attacks, including in the cross-domain case left open by prior work (Section~\ref{sec:bindmechref}). We use ciphertext-derived randomness modification for Bulletproofs (same-group) and a blinded arithmetic hash with embedded Schnorr constraint plus explicit-carry modular reduction for zk-SNARKs and zk-STARKs (cross-domain). Formal security reduces to discrete logarithm problem hardness and Minimal Multiplicative Complexity (MiMC) key-recovery under known plaintext (Appendix~\ref{app:stark-snark-proof}).

(2) A formal problem definition for secure decentralized aggregation with syntax and security properties, including anti-substitution binding (Section~\ref{sec:syntaxx}).

(3) Three complete Rust implementations (Appendix~\ref{app:artifact}) with production and defense-in-depth features, including rate limiting, fault tolerance, replay protection, and side-channel resistance (Sections~\ref{sec:bindmechref}, \ref{sec:security}).

(4) Demonstration and evaluation across 5--500 participants, including baseline comparison against unverified aggregation, quantifying the cost of verifiability at approximately 1.5\% overhead of a 10-second round or 0.25\% of a 60-second round (Section~\ref{sec:testing}).

\section{Problem Definition and Threat Model} \label{sec:syntaxx}

\subsection{System Model}
We consider $n$ participants $\mathcal{P}=\{P_1,...,P_n\}$. Each participant $P_i$ holds a private input $s_i \in \mathcal{S}$. The network shares a constraint predicate $\Phi:\mathcal{S} \to \{0,1\}$. An input is valid if $\Phi(s_i) = 0$. Honest participants hold valid inputs. A threshold  number of participants $t \le n$ determines the minimum quorum for decryption, where $n \ge 2t-1$ ensures an honest majority. The goal is to compute $\sum_{i=1}^{n}s_i$ without any party learning individual inputs, while guaranteeing that every $s_i$ in the aggregate satisfies $\Phi(s_i)=0$, with no trusted third party.

\subsection{Scheme Syntax}
A \textit{Secure Decentralized Aggregation} (SDA) scheme is a tuple of probabilistic polynomial-time (PPT) algorithms $\Pi = (\mathsf{Setup}, \mathsf{Enc}, \mathsf{Prove}, \mathsf{Vrfy}, \mathsf{Agg}, \mathsf{PDec}, \mathsf{Comb})$:

\begin{enumerate}
\item $\mathsf{Setup}(1^\lambda, n, t) \rightarrow (Y, \{x_i\}_{i \in [n]}, \mathsf{pp})$: Distributed key generation with security parameter $\lambda$. Produces shared public key $Y$, private key shares $\{x_i\}$, and public parameters $\mathsf{pp}$.

\item $\mathsf{Enc}(Y, s_i; r_i) \rightarrow c_i$: Encrypts input $s_i$ with randomness $r_i$ under public key $Y$, producing ciphertext $c_i$.

\item $\mathsf{Prove}(Y, s_i, r_i, c_i, \mathsf{pp}) \rightarrow \pi_i$: Produces a proof $\pi_i$ proving $\Phi(s_i) = 0$ and binding the proof to $c_i$.

\item $\mathsf{Vrfy}(Y, c_i, \pi_i, \mathsf{pp}) \rightarrow \{0,1\}$: Verifies that $\pi_i$ is both bound to $c_i$ and proves $\Phi(s_i) = 0$. Outputs 1 (accept) or 0 (reject).

\item $\mathsf{Agg}(\{c_i\}_{i \in \mathcal{V}}) \rightarrow \bar{c}$: Homomorphically aggregates ciphertexts from the set of participants $\mathcal{V}$ whose packages pass $\mathsf{Vrfy}$. Produces $\bar{c}$, an encryption of $\sum_{i\in\mathcal{V}} s_i$.

\item $\mathsf{PDec}(x_i, \bar{c}) \rightarrow (\sigma_i, \tau_i)$: Computes partial decryption $\sigma_i$ with correctness proof $\tau_i$.

\item $\mathsf{Comb}(\{(\sigma_i, \tau_i)\}_{i \in \mathcal{T}}, \bar{c}, \mathsf{pp}) \rightarrow A$: Combines at least $t$ partial decryptions with valid $\tau_i$ received from participants~$\mathcal{T}$. Decrypts $\bar{c}$ to produce $A = \sum_{i\in\mathcal{V}} s_i$.
\end{enumerate}

\begin{figure}[t]
    \centering
    \includegraphics[width=0.4\textwidth]{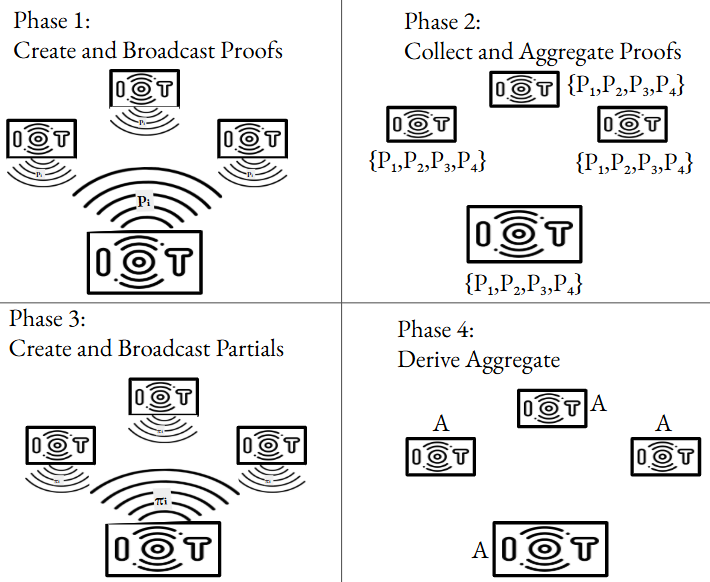}
    \caption{Architectural Diagram of zk-DEAP Protocol: In phases 1 and 2, participants encrypt and broadcast proof packages which are verified and aggregated. In phases 3 and 4, participants decrypt the aggregate by combining partial decryptions.}
    \label{fig:archi}
\end{figure}

Figure~\ref{fig:archi} shows how these algorithms fit together.

\subsection{Security Properties}

An SDA scheme $\Pi$ is \textit{secure} if it satisfies the following properties:

\noindent\textbf{Correctness.} If all parties follow the protocol, every party's  $\mathsf{Comb(\ldots)}$ outputs $\sum_{i=1}^{n} s_i$ with overwhelming probability.

\noindent\textbf{Privacy.} For any PPT adversary $\mathcal{A}$ controlling fewer than $t$ parties, and any two vectors of valid honest inputs $\vec{s}_0, \vec{s}_1$ with $\sum s_{0,i} = \sum s_{1,i}$, $\mathcal{A}$ cannot distinguish protocol executions on $\vec{s}_0$ from $\vec{s}_1$ with non-negligible advantage. Individual inputs are hidden; only the aggregate is revealed.

\noindent\textbf{Verifiability.} For any PPT adversary $\mathcal{A}$, there is negligible probability that $\mathcal{A}$ produces $(\pi, c)$ such that $\mathsf{Vrfy}(\pi, \ldots) = 1$ but $c$ encrypts some $s$ with $\Phi(s) \neq 0$.

\noindent\textbf{Binding.} For any PPT adversary $\mathcal{A}$, there is negligible probability that $\mathcal{A}$ produces $(\pi, c)$ such that $\mathsf{Vrfy}(\pi, \ldots) = 1$ but $\pi$ proves $\Phi(s) = 0$ while $c$ encrypts a different $s' \neq s$.

\noindent\textbf{Byzantine Tolerance.} If fewer than $t$ parties are corrupted, all honest parties compute the same correct aggregate, which includes every broadcast honest input and only valid inputs.

\noindent\textbf{Decentralization.} No single party's absence or corruption prevents honest parties from completing aggregation, provided at least $t$ honest parties participate.

\subsection{Adversary Model}
\label{sec:adversary-model}
We use the security parameter $\lambda = 128$, the standard industry target. We assume a PPT adversary $\mathcal{A}$ with polynomial-bounded parallelism and storage. The adversary can corrupt up to $t-1$ participants. The corrupted participants are defined as a subset of the network, $\mathcal{C} \subset \mathcal{P}$, where $|\mathcal{C}| < t$~\cite{canetti2001universally}. We also assume a static-corruption model where $\mathcal{C}$ is fixed before the protocol begins (chosen by $\mathcal{A}$ given public parameters but before DKG). Corrupted participants are immediately under adversarial control and remain so for the lifetime of the protocol. Corrupted participants can deviate arbitrarily from the protocol, coordinate attacks, and attempt to create malicious inputs, proofs, and commitments. The adversary can learn the entire internal state of the corrupted participants, including all proofs, secrets, and randomness history. Recent adaptive-security analyses of FROST~\cite{CritesKatzKomloTessaroZhu23,CritesKatzKomloTessaroZhu25} 
suggest our analysis could also extend to adaptive corruption, which we leave to future work.

We assume the adversary has visibility into the network and can intercept and read network traffic. We furthermore assume they can modify, inject, delay, replay, and reorder messages. We assume that although they may delay delivery or temporarily partition the network, all honest parties eventually receive the same set of proof packages, including all those broadcast by honest parties. The adversary cannot break cryptographic primitives or forge signatures without the respective secret keys. The adversary has access to the entire protocol specification, public parameters, historical transcripts, and all honest parties’ ciphertexts and proofs, including from past rounds. 

\subsection{Security Assumptions} Across all implementations, we assume the discrete logarithm problem (DLP) is hard for Ristretto255 based on the best known attack (Pollard's rho algorithm~\cite{pollard1978monte}), requiring expected $O(\sqrt{q}) \approx 2^{126}$ group operations. ElGamal, both traditional and lifted, provides indistinguishability under chosen-plaintext attack (IND-CPA) under Decisional Diffie-Hellman (DDH) hardness~\cite{tsiounis1998security, cramer1998practical}. Breaking DDH over Ristretto255 is believed to be computationally equivalent to solving DLP, requiring $\approx 2^{126}$ operations. We label this as 128-bit security per NIST convention~\cite{nist_sp800_186}, since the 2-bit gap from the conjectured best attack is within the standard NIST acceptability band. 

We model hash functions as random oracles. Under the Random Oracle Model (ROM), our $\Sigma$-protocols achieve the zero-knowledge property. Practical attacks against our transcript-based Fiat-Shamir instantiation require finding hash collisions (requiring $2^{128}$ queries for 256-bit outputs via birthday bound) or breaking underlying hash properties~\cite{bellare1993random, canetti2004random, fischlin2010random}. We also assume all participants have loose time synchronization within a skew $\Delta_{skew}$~\cite{gong1993timestamps}.

In terms of the specific ZKP implementations, 
Bulletproof soundness relies only on the DLP in Ristretto255, adding no new assumptions. zk-SNARKs using Halo2 assume q-Strong Diffie-Hellman (q-SDH) and pairing assumptions. zk-STARKs rely solely on collision-resistant hash functions, making them post-quantum secure. 
The default Winterfell $\mathbb{F}_p$ base field provides ${\sim}$100 bits of conjectured STARK proof soundness, and quadratic extensions to $\mathbb{F}_{p^2}$ achieve 128+ bits when required. This soundness level is independent of our binding mechanism's security, which provides full 128-bit hardness via the discrete logarithm assumption and MiMC key-recovery assumption. A STARK proof forgery would bypass the constraint $\Phi$ but cannot cause a substitution attack. Bulletproofs and STARKs operate transparently with no trusted setup, while Halo2 requires a Powers of Tau ceremony. 

\section{zk-DEAP Protocol Design}
\label{sec:bindmechref}

\subsection{Building Blocks}

\subsubsection{Ristretto255}
We use the Ristretto255 group, a prime order abstraction over Curve25519~\cite{pkc-2006-3351}, which has been specified by the IRTF in RFC 9496~\cite{rfc9496}. Ristretto255 underlies all our elliptic curve operations (ElGamal, Pedersen commitments, FROST, discrete log extraction), except the Halo2 variant which uses BN254 natively. We use the \texttt{dalek-ng} Rust library, which provides twisted Edwards point addition, NAF scalar multiplication, and 32-byte compression~\cite{curve25519-dalek-ng}.

\subsubsection{FROST}
We use the DKG protocol of FROST~\cite{KomloG20, RFC9591} to establish a distributed public key $Y$ where each participant holds a share of the private key $x_i$. Our implementation uses the standard 2-round interactive setup with the Rust \texttt{frost-ristretto255} library~\cite{zf_frost_ristretto255}.

\subsubsection{ElGamal}
We use lifted ElGamal HE to encrypt and aggregate the inputs so that no individual input is revealed. Lifted ElGamal encryption produces two-component ciphertexts $c_i = (c_1^{(i)}, c_2^{(i)}) = (g^{r_i}, g^{s_i} Y^{r_i})$ for Ristretto255 generator $g$, and our $\mathsf{Agg}$ algorithm multiplies ciphertexts component-wise, giving $(\bar{c}_1, \bar{c}_2) = (g^R, g^A Y^R)$ with $R = \sum_{i\in\mathcal{V}} r_i$.

Lifted ElGamal is a mature aggregation method, with recent scholarship even looking to make it adaptively secure~\cite{das2025threshold}. It has proven efficient in Blockchain and IoT environments~\cite{iot-blockchain-elgamal, iot-elgamal-wsan, baharon2024secure} compared to Paillier alternatives, which are based on RSA and have large sizes~\cite{paillier1999}, or Boneh-Goh-Nissim (BGN), which relies on heavier bilinear pairing operations~\cite{bgn2005}. 

We choose lifted ElGamal for three reasons. First, ciphertexts are 64 bytes (two Ristretto255 points), versus 512+ bytes for Paillier with 2048-bit moduli. Second, ElGamal operates natively on Ristretto255, sharing algebraic structure with FROST, Pedersen commitments, and the binding mechanism, where Paillier would require a separate framework for each component. Third, the Baby-Step Giant-Step (BSGS) extraction constraint is practical in most scenarios: where $n \leq 500$ participants with inputs in $[0,255]$, the maximum aggregate is $127{,}500$, requiring $\lceil\sqrt{127{,}500}\rceil = 358$ BSGS steps.

\subsubsection{Threshold Decryption and Extraction}
The challenge in combining FROST with ElGamal is decrypting the plaintext. 
The aggregate ciphertext $(\bar{c}_1, \bar{c}_2)$ is encrypted under the shared public key $Y$, so at least $t$ participants must submit partial decryptions $\sigma_i = \bar{c}_1^{x_i}$  using their share of the private key. As is standard for threshold ElGamal under DDH~\cite{cramer1997}, these can be combined using Lagrange coefficients to recover the plaintext $g^A=\bar{c}_2/\bar{c}_1^{x}$, where $A = \sum_{i\in\mathcal{V}} s_i$.

We can then extract the aggregate $A$ from $g^A$  by solving the discrete log. As is standard in the field, we can use the BSGS algorithm in $O(\sqrt{A_{\max}})$ time and space, where $A_{\max}$ is an upper bound on the aggregate (distinct from the participant count $|\mathcal{P}| = n$). This was first proposed by Daniel Shanks~\cite{shanks1971}, and has received many optimizations~\cite{stein-teske2005}. It has been specifically noted for its use in blockchains with ElGamal~\cite{chatzigiannis2021homomorphic}.

\subsection{Zero Knowledge Proofs}
Because our system's goal is zero-trust and inputs are aggregated while encrypted, we need a way for participants to verify conditions on these inputs without knowing what they are. This is the role of the ZKPs. We provide three implementations with three different ZKP systems for defining $\Phi$: Bulletproof, zk-SNARK, and zk-STARK. We first introduce the special-purpose $\Sigma$-protocols used by all three variants, which are unrelated to $\Phi$.

\subsubsection{$\Sigma$-Protocols} We use two non-interactive $\Sigma$-protocols~\cite{damgaard2010sigma,ciampi2016improved}: a Schnorr proof for discrete-log knowledge used to link the ElGamal ciphertext to the ZKP, and a Chaum-Pedersen proof for the discrete-log equality used in threshold decryption where each participant proves correct computation of $\sigma_i = \bar{c}_1^{x_i}$.

\subsubsection{Pedersen Commitments}
We use Pedersen commitments to bind inputs to the ZKPs~\cite{pedersen1991non}. Each participant $P_i$ uses the same randomness $r_i$ used for ElGamal encryption, and computes commitment $C(s_i,r_i)=g^{s_i}  h^{r_i}$, where $h$ denotes the appropriate second generator. For Bulletproofs we set $h = h_{BP}$, an independent generator with no known discrete log relation to $g$ or $Y$. For the SNARK/STARK variants we set $h = Y$ (the FROST distributed public key), which collapses the Pedersen commitment to the ElGamal $c_2$ component itself, so $C(s_i,r_i) = c_2^{(i)}$ by construction (a fact we exploit). This allows for perfect hiding and computational binding~\cite{damgaard2010sigma}.

\subsubsection{Bulletproofs}
Bulletproofs are our first option for defining $\Phi$. They are a type of non-interactive ZKP that specializes in range proofs, requires no trusted setup, and achieves logarithmic proof sizes~\cite{DBLP:conf/sp/BunzBBPCP18}. Built natively on top of Pedersen commitments, their core innovation lies in an efficient inner product argument that recursively reduces proof size from linear to logarithmic in the bit length $\ell$ of the range $[0,2^\ell)$. They also allow for linear proving and verification time. Bulletproofs have demonstrated effectiveness in blockchain-based IoT data sharing~\cite{li2024bulletproof-iot} and privacy-preserving energy applications~\cite{li2024energy-bulletproof}. 

\subsubsection{zk-SNARKs}

As an alternative to Bulletproofs, we also present an implementation using zk-SNARKs, specifically Zcash \texttt{halo2\_proofs}~\cite{bowe2021halo2}, a PLONK-based proof system~\cite{gabizon2019plonk} with KZG polynomial commitments~\cite{kate2010constant} operating over the BN254 pairing-friendly curve with scalar field $\mathbb{F}_r$ where $r \approx 2^{254}$. There are other popular variants, including Groth16, which offers smaller proofs with less flexible Rank-1 Constraint System (R1CS) constraints and complex per-circuit trusted setup communication~\cite{groth16}, or Plonky2, which operates over a Goldilocks field for fast proving without a trusted setup, but is still in development and requires complex, wide circuits and larger proofs~\cite{plonky2}.

\subsubsection{zk-STARKs}
Our last implementation uses zk-STARKs~\cite{ben2018scalable}, which rely on collision-resistant hash functions instead of elliptic curve cryptography, making them post-quantum sound. The Fast Reed-Solomon Interactive Oracle Proof of Proximity (FRI)~\cite{bensasson2019stark} probabilistically verifies polynomial degree bounds.  zk-STARKs need no trusted setup, as they derive all randomness publicly via Fiat–Shamir.

\subsection{Substitution Attacks}
\label{sec:attttack}

Having established the building blocks, we next describe the two substitution attack variants introduced in Section~\ref{section:intro} to motivate our binding mechanisms.

\textbf{Attack 1: Value Substitution.} An adversary generates $(s_i,r_i)$ with $\Phi(s_i)=0$ and produces a valid $\pi$ for $(s_i,r_i)$. The adversary then constructs $c'_2=g^{s'}Y^{r_i}$ for some $s'$ where $\Phi(s') \ne 0$, and submits $\pi$ with ciphertext $(c_1,c'_2)$. Without binding, $\mathsf{Vrfy}(\pi,...)=1$ because $\pi$ was honestly generated, but the ciphertext encrypts $s'$.

\textbf{Attack 2: Randomness Substitution.} Even if the ZKP is bound to $c_2$, a clever adversary can attempt to submit $(\pi, c'_1,c_2)$ where $c'_1=g^{r'}$ for $r' \ne r_i$. Since threshold decryption computes $\bar{c}_2/\bar{c}_1^{x}$, if $r' \neq r_i$ the randomness terms do not cancel and decryption fails or produces an incorrect aggregate. This is a denial-of-service attack undetectable by ZKP verification.

\subsection{Defense Strategy}

zk-DEAP performs ElGamal encryption under a FROST distributed public key to create an aggregate that can be decrypted only through the coordination of a threshold number of participants, enabling completely decentralized, secure aggregation while keeping individual inputs hidden. 
The protocol also enforces conditions on the ciphertexts using ZKPs, preventing adversaries from compromising the aggregate with malicious encrypted values.
Due to the disjoint nature of the ciphertexts and the proofs, a careful binding construction is necessary to avoid value substitution and randomness substitution attacks.

zk-DEAP therefore creates proof packages with three binding components, summarized here and described in detail for each ZKP variant in the following sections. To prevent value substitution attacks, we first use a randomness shift $\beta_i$ (Bulletproofs) or a pre-challenge commitment $V_\text{commit}$ (SNARK/STARK) to bind the ZKP to this ciphertext. We then use a Pedersen link (Equation (3), Bulletproofs) or an embedded Schnorr constraint (Equation (5), SNARK/STARK), which forces the value checked by the ZKP to equal the Schnorr-extracted $s_i$, so any $\Phi$ verified in the circuit applies to the encrypted value. To prevent randomness substitution attacks, a Schnorr link (Equations (1), (2) below) binds $(c_1^{(i)}, c_2^{(i)})$ to a single $(s_i, r_i)$ pair, preventing ciphertext component substitution. Ed25519 signatures seal the package against tampering.

\subsection{Bulletproof Binding Architecture} The Bulletproof implementation uses two distinct Pedersen commitment schemes with different generator pairs. After computing the ElGamal ciphertext using basepoint $g$ and public key $Y$, an ElGamal-Pedersen linking Schnorr proof commits to the same values using different generators, $(g,h_{BP})$: $C_{BP}(s_i,r_i)=g^{s_i} h_{BP}^{r_i}$. Using random nonces $w, v \overset{\$}{\leftarrow} \mathbb{Z}_q$, the prover computes three commitments $R = g^w$, $S = g^v Y^w$, and $P = g^v h_{BP}^w$, where the first two use ElGamal generators $(g,Y)$ and the third uses Bulletproof generators $(g,h_{BP})$. The Schnorr proof verifies:
\begin{align*}
\text{ElGamal $c_1$ correctness: } \quad g^{r_{resp}} = R  (c_1^{(i)})^ e\tag{1} \\
\text{ElGamal $c_2$ correctness: } \quad g^{s_{resp}}  Y^{r_{resp}} = S  (c_2^{(i)})^e \tag{2} \\
\text{Pedersen binding: } \quad g^{s_{resp}}  h_{BP}^{r_{resp}} = P  C_{BP}(s_i,r_i)^ e\tag{3}
\end{align*}
where $r_{resp}=w+e  r_i$ and $s_{resp}=v+e s_i$ are the Schnorr responses and $e$ is the Fiat-Shamir challenge. The binding factor $\beta_i$ does not appear in Equations (1)-(3); it is applied separately at the Bulletproof verification step, against the adjusted commitment $C'_{BP}$ defined below. This adjustment proves the ElGamal ciphertext (using generators $g$,$Y$) and the Bulletproof commitment (using generators $g,h_{BP}$) both encrypt/commit to the same input $s_i$ and randomness $r_i$ despite the different algebraic bases. To bind the Bulletproof to the ElGamal ciphertext, we derive a binding factor:
\[
\beta_i= \mathcal{H}_{FS}(\mathbb{S},c_1^{(i)},c_2^{(i)},id_i,ts_i,nonce_i)
\]
using the Merlin transcript system~\cite{merlin2018}, which is built on the STROBE framework~\cite{hamburg2017strobe} over Keccak-$f$[1600]~\cite{bertoni2013keccak}, where $\mathbb{S}$ is a string for domain separation, and $id_i,ts_i,nonce_i$ correspond to the participant ID, timestamp, and nonce of the proof. The Bulletproof is then generated with modified randomness $r'_i=r_i+\beta_i$, so during verification, peers reconstruct $\beta_i$ and verify the Bulletproof against the adjusted commitment:
\[
C'_{BP}(s_i,r'_i)=C_{BP}(s_i,r_i)  h_{BP}^{\beta_i}=g^{s_i}  h_{BP}^{r_i+\beta_i}
\]

The separation of the two commitment schemes allow the ElGamal and Bulletproof to use their respective generator bases while maintaining a cryptographic link between them. Any ciphertext swap (Attack 1) requires a hash collision or breaking commitment binding.

\subsection{SNARK/STARK Binding Architecture}
Unlike Bulletproofs, where we can modify the commitment randomness directly due to their native Pedersen commitment structure, zk-SNARKs and zk-STARKs operate over different algebraic structures. The core challenge is ensuring that the ZKP's private witness $s_\text{ZKP}$ equals the value $s_i$ encrypted in the ciphertext, although they live in different algebraic domains (Ristretto255 for ElGamal, $\mathbb{F}_p$ for the ZKP).

As established, this is an example of the cross-domain binding problem: we must connect a proof system's private witness to a commitment in a different algebraic structure without revealing the witness or requiring the proof system to simulate the commitment's group arithmetic. Commit-and-prove SNARKs (LegoSNARK~\cite{campanelli2019legosnark}, Eclipse~\cite{aranha2022eclipse}, SAVER~\cite{lee2019saver}) provide linking when the commitment and proof system share a field or are related by a known pairing. Cross-prime linking ($\mathbb{Z}_q \to \mathbb{F}_p$ with $p, q$ distinct primes, no embedding) without highly inefficient non-native circuit arithmetic remains an open problem. Our solution avoids both. We embed a Schnorr linear constraint directly within the ZKP circuit, combined with a blinded, key-recovery-hard arithmetic hash that commits to the Schnorr nonce before the Fiat-Shamir challenge. This ensures that any constraint $\Phi$ verified by the circuit on $s_\text{ZKP}$ automatically applies to the encrypted value $s_i$, without crossing algebraic domains or sacrificing privacy.
 
\subsubsection{Construction} We use shared generators $(g, Y)$ for both ElGamal and Pedersen commitments, yielding $c_2^{(i)} = g^{s_i}  Y^{r_i} = C(s_i, r_i)$ by construction. The ElGamal-Pedersen linking Schnorr proof equations (1) and (2) establish ciphertext well-formedness with response $s_\text{resp} = v + e  s_i$, where $v$ is the Schnorr nonce committed before challenge derivation.
 
The binding mechanism then operates as follows:

(1) The prover computes $v_f = v \bmod p$ and samples a random blinding factor $b \overset{\$}{\leftarrow} \mathbb{F}_p$.

(2) The prover computes \mbox{$V_\text{commit}=\text{MiMC}_{63}(v_f;\; s_\text{ZKP}, \epsilon_i, b)$,} an iterated MiMC hash~\cite{cryptoeprint:2016/492} using $v_f$ as the initial state, with $(s_\text{ZKP}, \epsilon_i, b)$ mixed in as round keys, where $\epsilon_i = \varphi(C(s_i, r_i))$ encodes the Pedersen commitment as a field element. For an honest prover, $s_{\text{ZKP}} = s_i$; the binding argument uses $s_{\text{ZKP}}$ since the circuit enforces the MiMC relation on this witness, which the binding mechanism then forces to equal $s_i$. The encoding $\varphi$ maps a compressed Ristretto255 point (32 bytes) to $\mathbb{F}_p$ by XOR-folding the upper and lower 16-byte halves. Modeling $\varphi$'s image heuristically as uniform over $\mathbb{F}_p$ (a standard approximation, tight up to the $\sim 1/8$ density of valid Ristretto encodings in the byte space, a constant factor absorbed into the bounds), two distinct Ristretto points map to the same $\mathbb{F}_p$ element with probability $\sim 1/p \approx 2^{-128}$ per pair, and a birthday search for any colliding pair costs $\sim 2^{64}$ work. Importantly, $\varphi$'s collision profile is not load-bearing for anti-substitution. Binding rests on two independent mechanisms: the Schnorr link (Equations (1) and (2)) binds $(c_1^{(i)}, c_2^{(i)})$ to the unique $(s_i, r_i)$ pair the prover knows, so any ciphertext swap breaks Schnorr extraction, and the embedded Schnorr constraint (Equation (5)) forces $s_\text{ZKP}$ to equal that same value. $\epsilon_i$ is a third, defense-in-depth layer preventing transplantation across distinct ciphertexts that encrypt the same plaintext. Overall binding reduces to DLP hardness and MiMC key-recovery assumption, not to $\varphi$'s collision resistance.

(3) $V_\text{commit}$ is appended to the Merlin transcript before the Fiat-Shamir challenge $e$ is derived.

(4) After challenge derivation, the prover computes $\text{resp}_f = s_\text{resp} \bmod p$ and $e_f = e \bmod p$.

(5) The ZKP (SNARK or STARK) proves, using $s_\text{ZKP}$, $v_f$, $b$, and $\omega$ as private witnesses:
\setcounter{equation}{3}
  \begin{align}
    \Phi(s_\text{ZKP}) &= 0 \label{eq:phi} \\
    \text{resp}_f &= v_f + e_f  s_\text{ZKP} - \omega  \tilde{q} \label{eq:schnorr_linear} \\
    V_\text{commit} &= \text{MiMC}_{63}(v_f;\; s_\text{ZKP}, \epsilon_i, b) \label{eq:mimc}
  \end{align}
 
The public inputs to the ZKP are $V_\text{commit}$, $\text{resp}_f$, $e_f$, and $\epsilon_i$. The private witnesses are $s_\text{ZKP}$, $v_f$, $b$, and the carry value $\omega$. The term $\tilde{q} := q \bmod p$ (the Ristretto group order reduced into the proof system's field) and $\omega$ together handle the modular reduction between $\mathbb{Z}_q$ (Schnorr) and $\mathbb{F}_p$ (proof system). The prover sets $\omega = \lfloor (v + e  s_\text{ZKP}) / q \rfloor$, the integer number of wraps of the Schnorr sum, and the circuit (or, for STARKs, the Algebraic Intermediate Representation, AIR) enforces $\omega \in [0, S_\text{max}]$ via an in-proof range-check, where $S_\text{max} = \max_{s \in \mathcal{S}_\Phi} s$ is the input bound implied by $\Phi$. The carry width thus tracks the input domain: $\omega \in \{0,1\}$ (a single bit) for our implemented binary constraint $\mathcal{S}_\Phi = \{0,1\}$ (explained below), and $\lceil \log_2(S_\text{max}{+}1) \rceil$ bits for any bounded domain. The binding reduction (Appendix~\ref{app:stark-snark-proof}) is identical regardless of width: once $\omega$ is correctly reduced, Equation (5) forces $s_\text{ZKP} = s_\text{schnorr}$ and Equation (6) forces $v_f$ to be the true Schnorr nonce, so any $\Phi$ enforced in the circuit applies to the encrypted value. Both the SNARK and STARK variants use the same explicit-$\omega$ construction: $\omega$ is a private witness, Equation (5) is enforced as a linear relation in witnesses, and $\omega \in [0, S_\text{max}]$ is range-checked within the circuit/AIR. The reduction below applies uniformly to both.
 
\subsubsection{Binding} Equation (5) forces $s_\text{ZKP}$ to equal the value $s_i$ used in the Schnorr proof, because the Schnorr nonce $v$ is committed before the challenge via $V_\text{commit}$. Specifically, an adversary encrypting $s' \neq s_\text{ZKP}$ (with $s_\text{schnorr} = s'$) obtains $\text{resp}_f = v_f + e_f  s' - \omega \tilde{q}$, so Equation (5) forces $v_f' = v_f + e_f(s' - s_\text{ZKP}) + (\omega' - \omega)\tilde{q} \neq v_f$ (since $e_f$ is a random oracle output and $s' \neq s_\text{ZKP}$, the $e_f(s' - s_\text{ZKP})$ term dominates). Since $V_\text{commit} = \text{MiMC}_{63}(v_f;\; s_\text{ZKP}, \epsilon_i, b)$ was committed before $e$ was known, satisfying Equation (6) with the shifted $v_f'$ requires producing $\text{MiMC}_{63}(v_f';\; s_\text{ZKP}, \epsilon_i, b') = V_\text{commit}$ for some $b'$. Since MiMC with fixed round keys is a permutation on $\mathbb{F}_p$ (as $x^5$ is a bijection when $\gcd(5, p{-}1) = 1$), different initial states map to different outputs under the same key schedule. The adversary must therefore recover round-key material $b'$ such that, under the protocol-fixed initial state $v_f'$, MiMC produces the same output. This is the MiMC key-recovery-under-known-plaintext game, $2^{128}$-hard with best known attacks at 63 rounds, and is within the well-studied cryptanalytic regime for $x^5$-MiMC. See Appendix~\ref{app:stark-snark-proof} for the formal reduction.
 
\subsubsection{Privacy} The verifier sees $V_\text{commit}$, $\text{resp}_f$, $e_f$, and $\epsilon_i$, but not $s_\text{ZKP}$, $v_f$, $b$, or $\omega$. From Equation (5), $\text{resp}_f = v_f + e_f  s_\text{ZKP} - \omega \tilde{q}$ has three unknowns and provides standard Schnorr zero-knowledge: any candidate $s$ yields a candidate $v_f(s) = \text{resp}_f - e_f  s + \omega \tilde{q}$. To verify a guess against $V_\text{commit}$ the verifier would need the blinding factor $b$. Since $b$ is uniform on $\mathbb{F}_p$ and unrevealed, recovering it from $V_\text{commit}$ is exactly the MiMC key-recovery problem on which binding rests, requiring $\sim 2^{128}$ work. Individual inputs are therefore computationally hidden, and only the aggregate is revealed through threshold decryption. (Note: $v_f = v \bmod p$ where $v$ is uniform on $\mathbb{Z}_q$; the resulting statistical distance from uniform on $\mathbb{F}_p$ is $O(p/q) \approx 2^{-124}$, negligible and absorbed into the bound.)
 
\subsubsection{Anti-Substitution} The commitment encoding $\epsilon_i = \varphi(C(s_i, r_i))$ is a public input to the ZKP. Changing the ciphertext changes $\epsilon_i$, invalidating the MiMC computation and requiring a completely new ZKP proof. This prevents proof transplantation across ciphertexts.
 
\subsubsection{MiMC Construction} MiMC iterates the round function $F(x) = (x + k_i)^5$ over $\mathbb{F}_p$, where $k_i = s_\text{ZKP} + \epsilon_i + b + \mathit{rc}_i$ combines the private round keys with public round constants $\mathit{rc}_i$ derived deterministically from BLAKE3 labels. We use 63 rounds with the $x^5$ S-box, matching a 64-row STARK trace (one round per transition). The MiMC paper recommends $\lceil 128 / \log_2 5 \rceil = 56$ rounds for $|\mathbb{F}_p| \approx 2^{128}$, so our 63 rounds provide a comfortable security margin. The $x^5$ S-box is a bijection on $\mathbb{F}_p$ when $\gcd(5, p{-}1) = 1$, which holds for the Winterfell prime $p = 2^{128} - 45 \cdot 2^{40} + 1$. We use MiMC because of its well-established efficiency and security within ZKPs. The main other mature alternative is Poseidon~\cite{cryptoeprint:2019/458}, which uses the same $x^5$ S-box design, but is unnecessarily complex and inflexible given our use case.
 
Together, these form the binding architecture:
\[
\resizebox{\columnwidth}{!}{$
\Phi(s_{\text{ZKP}}) \xleftarrow[\text{constraint}]{s_{\text{ZKP}}}
\bigl(v_f + e_f s_{\text{ZKP}} - \omega \tilde{q} = \text{resp}_f\bigr)
\xrightarrow[\text{pre-challenge}]{\text{MiMC}(v_f; s, \epsilon, b)}
V_{\text{commit}}
$}
\]
Any constraint $\Phi$ added to the circuit on $s_\text{ZKP}$ automatically applies to the encrypted value, because the embedded Schnorr constraint forces $s_\text{ZKP} = s_i$.

The dual-generator approach employed in our Bulletproof implementation generally adheres to established practices in proof composition~\cite{bunz2020zether}. The blinded MiMC binding in the SNARK/STARK versions is novel, and draws from advances in arithmetic hash functions designed for zero-knowledge proof systems~\cite{cryptoeprint:2016/492, cryptoeprint:2019/458}, combined with techniques for cross-domain proof linking~\cite{cryptoeprint:2019/1021}. As such, we provide a formal security proof in Appendix~\ref{app:stark-snark-proof}.

\subsubsection{ZKP Transcripts} All protocols use Merlin transcripts~\cite{merlin2018} for domain separation, deterministic Fiat-Shamir, and binding of proof components to execution context (timestamps, IDs, nonces). The Bulletproof variant adds two transcripts for range proof binding and binding factor derivation.

\subsection{ZKP-Specific Implementations}

\subsubsection{Bulletproof Implementation} We use a dual-generator architecture with ElGamal operating over $(g,Y)$ and the Bulletproof operating over $(g,h_{BP})$. This separation prevents attacks where knowledge of the discrete logarithm relationships compromises soundness. Unlike the zk-SNARK and zk-STARK implementations, Bulletproofs are already integrated with Pedersen commitments and thus use randomness modification rather than a hash constraint for binding. For the input constraint, we use an 8-bit range proof from the Dalek \texttt{bulletproofs} crate~\cite{dalek-bulletproofs}. The 8-bit range constraint demonstrates bounded contribution verification, ensuring inputs fall within valid ranges, which addresses problems such as gradient poisoning in FL deployments.

\subsubsection{zk-SNARK Implementation} 

Our Halo2 circuit enforces three types of constraints, structurally identical to the STARK variant. First, a blinded arithmetic hash commitment $V_\text{commit}$ realized as an iterated MiMC$_{63}$ chain over $\mathbb{F}_r$ (BN254's scalar field) binds the circuit to the Schnorr proof via the nonce $v_f$. Second, the Schnorr binding constraint $\text{resp}_f = v_f + e_f  s_\text{ZKP} - \omega  \tilde{q}$ with $\omega$ as a private witness and $\omega \in \{0,1\}$ enforced by an in-circuit range-check gate. This forces $s_\text{ZKP}$ to equal the encrypted value. Third, the input constraint. For our demonstration, we employ a binary constraint $s_{\text{ZKP}} \in \{0, 1\}$ via $s_{\text{ZKP}}  (s_{\text{ZKP}} - 1) = 0$. 
Because $s_\text{ZKP}$ is bound to the ciphertext, any constraint on $s_\text{ZKP}$ automatically applies to the encrypted value. In a federated learning deployment, this could address free-rider attacks where participants benefit from gradient updates without contributing~\cite{lin2019freeridersfederatedlearningattacks}. More complex constraints (such as gradient norm bounds or cosine-similarity checks) can be substituted without modifying the binding mechanism. It is important to note that an insecure or loose choice of $\Phi$ can still break security by allowing adversaries to contribute malicious inputs that still satisfy $\Phi$.

zk-SNARKs require a trusted setup. We derive parameters from the PSE/Hermez Perpetual Powers of Tau ceremony~\cite{perpetual-powers-tau} ($k=8$), which is secure provided one participant destroys their share~\cite{bowe2017scalable}. Participants receive these public parameters at setup, which can be verified against published hashes to detect compromised versions. 

\subsubsection{zk-STARK Implementation} 
We use Facebook's \texttt{winterfell} library~\cite{winterfell-facebook}, operating over the prime field $\mathbb{F}_p$ where $p = 2^{128} - 45 \cdot 2^{40} + 1$. Like zk-SNARKs, the STARK enforces the binding mechanism and input constraint within the AIR. The execution trace is a matrix $\mathcal{T} \in \mathbb{F}^{\mathfrak{n} \times \mathfrak{m}}$. In our implementation, $\mathfrak{m} = 6$ columns represent input ($s_\text{ZKP}$), MiMC state (evolving from $v_f$ to $V_\text{commit}$), commitment encoding ($\epsilon_i$), blinding factor ($b$), a Schnorr binding auxiliary, and the carry value $\omega$, across $\mathfrak{n} = 64$ rows with 63 transitions.

The AIR defines eight transition constraints. The transition constraints are the input gate $\mathcal{T}[0,i](\mathcal{T}[0,i]-1)=0$, the MiMC round $\mathcal{T}[1,i{+}1] = (\mathcal{T}[1,i]+\mathcal{T}[0,i]+\mathcal{T}[2,i]+\mathcal{T}[3,i]+\mathit{rc}_i)^5$, constancy of $\mathcal{T}[0],\mathcal{T}[2],\mathcal{T}[3],\mathcal{T}[5]$, the Schnorr auxiliary $\mathcal{T}[4,i] = \mathcal{T}[1,i]+e_f\mathcal{T}[0,i]-\mathcal{T}[5,i]\tilde{q}$ with $\tilde{q}=q\bmod p$, and the carry range-check $\mathcal{T}[5,i](\mathcal{T}[5,i]-1)=0$. Boundary assertions $\mathcal{T}[1,63]=V_\text{commit}$, $\mathcal{T}[2,0]=\epsilon_i$, $\mathcal{T}[4,0]=\text{resp}_f$ tie the trace to public inputs $(V_\text{commit},\epsilon_i,\text{resp}_f,e_f)$. $\omega\in[0,S_\text{max}]$ is enforced within the AIR as a private witness ($\{0,1\}$ for our binary case). Columns 0, 1, 3, 5 remain private.

\subsection{High Level Overview} 
\begin{figure}[t]
    \centering
    \includegraphics[width=0.5\textwidth]{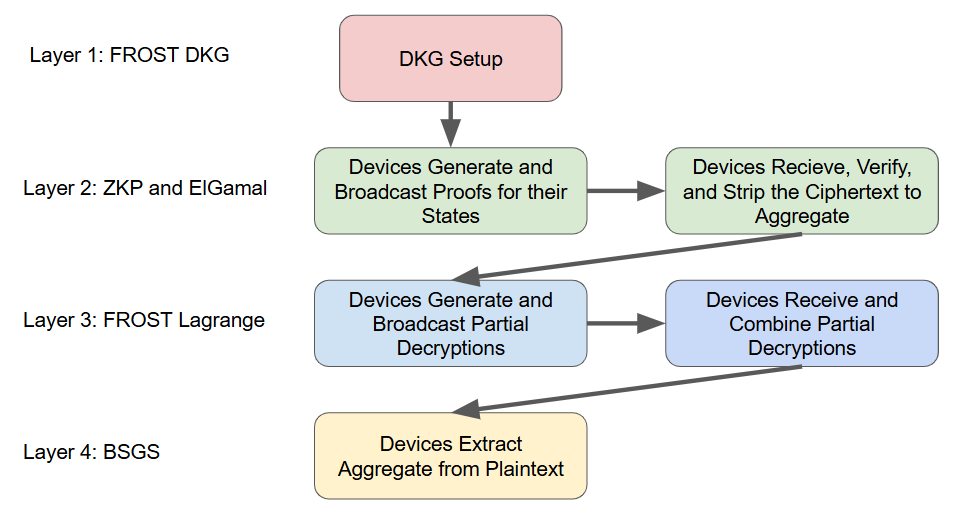}
    \caption{zk-DEAP Flowchart.}
    \label{fig:flowchart}
\end{figure}

We now summarize how the components described in the previous sections integrate into a complete system, as illustrated in Figure~\ref{fig:flowchart}.

\subsubsection{Setup and Proof Generation/Reception} 
\begin{figure}[t]
    \centering
    \includegraphics[width=0.45\textwidth]{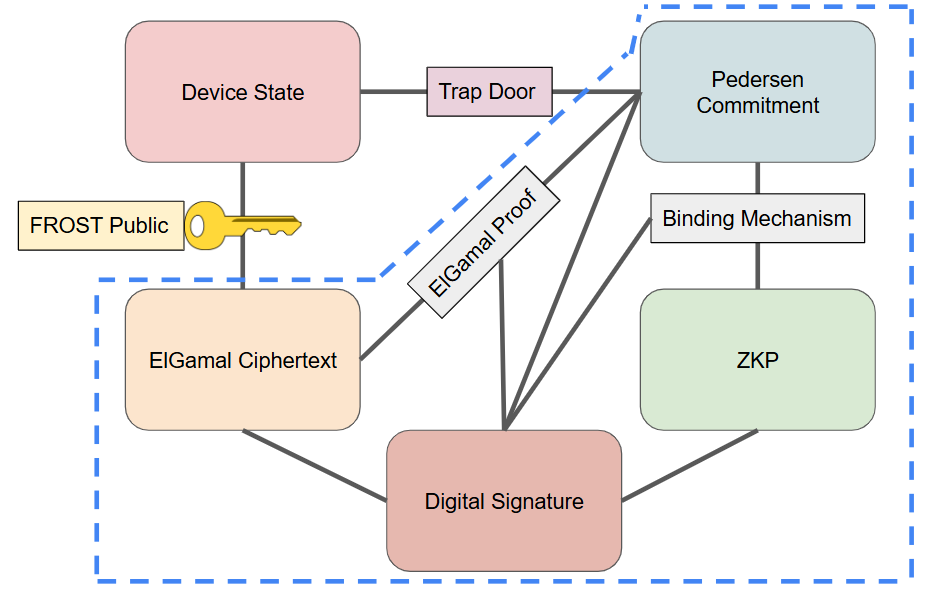}
    \caption{Diagram of Proof Package: The input is encrypted to produce an ElGamal ciphertext and bound to a Pedersen Commitment, which the ZKP enforces conditions on. An ElGamal Schnorr Proof binds the ciphertext to the commitment, which a binding mechanism binds to a ZKP. A digital signature then signs all the elements, and the participant's input is hidden.}
    \label{fig:proof-structure}
\end{figure}

After FROST DKG establishes the shared public key and secret shares, and participants exchange Ed25519 verification keys (with Halo2 participants also receiving Powers of Tau parameters), each participant encrypts its input under ElGamal and constructs a proof package (Figure~\ref{fig:proof-structure}).

Upon receiving a proof package, each participant first performs lightweight checks and verifies the Ed25519 signature. Next, it validates the Schnorr proof linking the ciphertext to the commitment, and the ZKP for constraint satisfaction. Finally, it homomorphically aggregates the ciphertext with previously verified ciphertexts, without knowing the individual input. 

\subsubsection{Threshold Decryption and Extraction} To decrypt the aggregated value, each participant generates a partial decryption using its secret share and a Chaum-Pedersen proof demonstrating correct computations, which are broadcast. These partial decryptions are received, and the Chaum-Pedersen proofs are verified. After collecting at least $t$ valid partial decryptions, each participant combines them using Lagrange interpolation, and uses the BSGS algorithm to recover the aggregate value from the plaintext. 

\section{Security Analysis}
\label{sec:security}

\subsection{Binding Security}

\begin{theorem}[Binding Security]
Under the discrete logarithm assumption on Ristretto255 and the $\mathrm{MiMC}_{63}$ key-recovery assumption (formally stated in Appendix~\ref{app:stark-snark-proof}), no PPT adversary can produce a valid proof package $(\pi, c_1, c_2)$ where $(c_1, c_2)$ encrypts $s$ with $\Phi(s) \neq 0$, except with negligible probability.
\end{theorem}

\begin{proof}[Proof sketch] From the Schnorr proof's special soundness, we extract $(s_\text{schnorr}, r)$ such that $c_1 = g^r$ and $c_2 = g^{s_\text{schnorr}}  Y^r$. The ZKP's knowledge soundness extracts $s_\text{ZKP}$, $v_f$, $b$, $\omega$ satisfying Equations (4-6). If $s_\text{ZKP} \neq s_\text{schnorr}$, set $\delta = s_\text{schnorr} - s_\text{ZKP} \neq 0$ and Equation (5) forces $v_f' = v_f + e_f  \delta + (\omega' - \omega)  \tilde{q}$. Satisfying Equation (6) with this shifted $v_f'$ either requires a carry collision (probability $\le q_H(S_\text{max}+1)/p$, negligible) or a MiMC key-recovery, contradicting the assumption. Thus $s_\text{ZKP} = s_\text{schnorr} = s$, and $\Phi(s) = 0$ by Equation~(4). Modular reduction is handled uniformly in both variants by the explicit carry value $\omega \in [0, S_\text{max}]$ enforced in the circuit/AIR. Both constructions achieve binding for arbitrary $\Phi$. See Appendix~\ref{app:stark-snark-proof} for the full proof and concrete bound.
\end{proof}

\begin{corollary}[Ciphertext Consistency]
The Schnorr extraction yields $r$ such that $c_1 = g^r$ and $c_2 = g^s  Y^r$. Any $c_1' \neq g^r$ fails Equation (1) verification, preventing randomness substitution (Attack 2).
\end{corollary}

\subsection{Shared Attacks Mitigated}
\label{sec:shared-attacks}
As we have demonstrated throughout the paper, the protocol's key security features derive from its privacy-preserving, decentralized, zero-knowledge proof design. The use of a cryptographically bound combination of ElGamal, FROST, and ZKPs, alongside signatures and defense-in-depth (DiD) features, prevents individual input disclosure while maintaining the security of the aggregate with no trust relationships. We validate these guarantees with Rust test demonstrations that simulate each attack vector, available in our open-source repository. Beyond the core cryptography, we also include production and DiD features, the most important of which are enumerated below.

\subsubsection{Byzantine Security} Our design logic centers around consensus and distributed system attacks. We prevent single point of failure (SPOF) and key escrow considerations by operating in a completely decentralized environment. Our protocol achieves Byzantine fault tolerance under the honest majority assumption~\cite{lamport1982byzantine, cachin2011introduction}, maintaining correctness and liveness provided: (1) Safety: if $|C| < t$ (fewer than $t$ participants are Byzantine), all honest participants compute identical aggregate values, (2) Liveness: at least $t$ honest participants participating in threshold decryption ensures protocol termination, and (3) Agreement: all honest participants verify the same ciphertexts and compute the same encrypted aggregate through deterministic homomorphic aggregation, given the delivery assumption in Section~\ref{sec:adversary-model}. The Chaum-Pedersen proofs enable $O(1)$ verification time per partial decryption, allowing immediate identification of Byzantine participants.

Byzantine attacks, including Sybil and eclipse attacks, are mitigated through a fixed participant set after DKG, and a threshold ensuring no subset can monopolize the protocol. Redundancy in the threshold can prevent censorship and targeted exclusion, while the deterministic recompute mitigates fork attacks. Liveness and ordering attacks are prevented with timeout mechanisms and commutative aggregation. The Chaum-Pedersen proof structure and threshold decryption prevent malicious and forged partial submissions, share interpolation attacks, and rushing attacks on partials.

The threshold architecture provides three distinct resilience properties, each with its own bound. Privacy holds against up to $t-1$ corrupted participants. No coalition smaller than the threshold can reconstruct the secret key or decrypt individual ciphertexts. Liveness tolerates up to $n-t$ crashes or non-participations, since at least $t$ honest partial decryptions must be collected to terminate. Aggregate correctness is independent of both bounds. Every submitted proof is individually verified, so any number of Byzantine participants producing invalid proofs is filtered before aggregation. The honest-majority constraint $n \geq 2t-1$ ensures privacy and liveness can hold simultaneously. Additional honest participants beyond the threshold provide redundancy~\cite{castro1999practical}, with potential for load balancing with additional coordination.

\subsubsection{Side-Channel Resistance} All sensitive cryptographic material is explicitly zeroized after use via the \texttt{zeroize} crate~\cite{zeroize2024}. We also implement a hardened implementation of the BSGS discrete log extraction algorithm using the \texttt{subtle}~\cite{subtle2024} crate to prevent timing side-channels~\cite{kocher1996timing}.

\subsubsection{DoS and DDoS Protections} To mitigate DoS and DDoS attacks, we use a multi-layered approach. First, we employ customized rate limiting per-participant and globally, using maximum messages within a fixed window and a maximum total rate. We also impose resource bounds where proofs are capped at a certain size depending on the implementation, storage is bounded for proofs and partials, and each participant is only allowed a certain number of nonces to protect against memory exhaustion. We then use early verification and lightweight checks before expensive cryptographic operations, rejecting proofs based on invalid signatures, nonces, timestamps, or sizes. Finally, automatic cleanup is employed to purge stale proofs and prune the rate-limiting input.

\subsubsection{Replay Attacks} To mitigate replay attacks, we employ nonce tracking, where each participant maintains a set of observed nonces per peer, and duplicate nonces are rejected. Combined with participant-specific ID tracking, this prevents traditional replays, which resubmit old messages, and duplication attacks, which submit the same proof multiple times~\cite{syverson2001replay}. We also employ timestamp validation with a skew tolerance~\cite{mills1991internet} to prevent proofs with future timestamps and stale proofs. This prevents delayed replay attacks where adversaries store valid proofs and resubmit them later. Thus, replay resistance is achieved through timestamps, nonces, and participant IDs. Even if an adversary captures a valid proof, replay attempts fail because (1) timestamps fail expiry checks, (2) nonces are detected as duplicates, and (3) digital signatures fail verification if any component changes.

\subsubsection{Authentication and EdDSA} All messages are authenticated using Ed25519 signatures~\cite{bernstein2012highs} in a strict verification mode that rejects non-canonical $s$ encodings, small-order $R$ points, and cofactor-equivalent signatures. This guarantees that any change to any signed component causes verification to fail. Each participant also keeps a valid participant set, which is established during the DKG phase, preventing Sybil attacks where adversaries create decentralized fake identities to influence the protocol~\cite{douceur2002sybil}.

\subsection{Security Limitations} 
\label{sec:security-lims}

We identify the following security limitations, which would not prevent use in common applications.

\subsubsection{Arithmetic Hash Security}
The binding mechanism for zk-SNARKs and zk-STARKs relies on MiMC key-recovery under known plaintext: given a single plaintext-ciphertext pair $(v_f, V_\text{commit})$ produced under an unknown blinding key $b$, no efficient adversary can recover an equivalent key $b'$ producing the same image. ($v_f$ is given to the adversary, not chosen, since the protocol fixes it via the post-challenge Schnorr-linear constraint.) For an ideal random permutation, this key recovery is $2^{128}$-hard, and best known attacks against \mbox{$x^5$-MiMC} remain above this bound at 63 rounds. While these arithmetic hash functions have seen deployment in production systems and withstood extensive cryptanalysis, they are less mature than traditional hash functions such as SHA-256 or BLAKE2b. The 63-round MiMC construction with $x^5$ S-box over $\mathbb{F}_p$ ($|p| \approx 2^{128}$) exceeds the recommended minimum of $\lceil 128/\log_2 5 \rceil = 56$ rounds. The Bulletproof variant is not affected by this consideration.

\subsubsection{STARK Proof Soundness} The Winterfell $\mathbb{F}_p$ base field ($|p| \approx 2^{128}$) with no extensions, which we used in our evaluation, provides approximately 100 bits of conjectured STARK proof soundness~\cite{winterfell-facebook}. This is independent of the binding mechanism's security, which provides 128-bit security via MiMC key-recovery under known plaintext and DLP hardness. For deployments requiring full 128-bit STARK soundness, Winterfell supports quadratic field extensions ($\mathbb{F}_{p^2}$), which achieve 128+ bit soundness at the cost of approximately 50\% larger proofs and increased proving time, as the additional evaluations and FRI layer require extension multiplication, which costs three base multiplications. The Bulletproof and SNARK variants are not affected by this consideration.

\subsubsection{Threshold Corruption and DKG Compromise} The fundamental attack that is not mitigated is compromise of a threshold number of participants, which would allow the adversary to decrypt any individual input. Furthermore, regardless of the threshold, an adversary who can corrupt a sufficient number of participants before or after the DKG setup can influence the aggregate using legitimate operations~\cite{GennaroJareckiKrawczykRabin07}. This is why selecting a good threshold, depending on the requirements of the situation, is crucial.

\subsubsection{Infrastructure and Deployment Risks} In production, without additional case-specific mitigation, the system is also vulnerable to total network isolation, global clock failure, advanced DoS/DDoS attacks, and sophisticated side-channels. The zk-SNARK implementation is vulnerable if corrupted public parameters are provided to the participants at startup~\cite{bowe2017scalable}.

\subsubsection{Forward Secrecy} Our protocol, as it stands, does not provide forward secrecy. If the FROST distributed private key (requiring $t$ shares) is compromised post-protocol, all historical ciphertexts using that key become decryptable. In environments where this is a concern, the DKG setup can be run on a schedule, rather than just once at the start, with retired shares zeroized. The specific schedule would depend on the application, and is outside the scope of this paper, but an important future direction.

\section{Testing and Efficiency}
\label{sec:testing}

\subsection{Testing Methodology}

We evaluate zk-DEAP's cryptographic performance directly, measuring each protocol operation in isolation rather than embedding the protocol within a specific application such as federated learning. This methodology is consistent with established practice for cryptographic protocol papers at NDSS and peer venues. For example, the NDSS 2024 threshold ECDSA evaluation~\cite{threshold-ecdsa-ndss2024} benchmarks signing and key generation operations, the NDSS 2026 distributed SNARK paper Cirrus~\cite{cirrus-ndss2026} measures proving time, and the NDSS 2026 cwPSU paper~\cite{cwpsu-ndss2026} reports communication and runtime for set operations. Evaluating the cryptographic layer independently also supports our general-purpose design claim, as the same measurements apply whether zk-DEAP is deployed for federated learning gradient aggregation, smart grid telemetry, or electronic voting. We do, however, contextualize these measurements against federated learning round times to demonstrate practical deployability.

We run the protocol on a 12th Gen Intel Core i7-1255U with 16 GB of RAM inside a Docker image using Rust 1.97 on Debian bookworm, with cryptographic operations restricted to a single thread to establish a baseline for parallelization. A fixture server was set up to pre-compute deterministic artifacts to ensure minimal latency when recording the operations. Next, a comprehensive test harness was run with 30 tests for nine key algorithms: DKG1, DKG2, DKG3, aggregate computation, proof generation, proof reception, partial generation, partial reception, and an unverified baseline. Additional tests were run for network scaling functions, covering participant sets from 5 to 500 and four different threshold options (0.25, 0.5, 0.67, 0.75). We measured wall time, CPU time, and artifact output sizes.

\subsection{ZKP Comparisons}

There are several important distinctions between the Bulletproof, zk-SNARK, and zk-STARK variants, as shown in Table~\ref{tab:zkp-compare}. It is first important to note, however, that a comparison between Bulletproofs and SNARKs/STARKs is not apples-to-apples, because of the different machinery that comprises them. The Bulletproof is an 8-bit range proof over a Pedersen commitment, with binding folded into the commitment randomness. The SNARKs and STARKs encompass a larger statement, with a 63-round MiMC chain, a Schnorr linear constraint, a range check on $\omega$, and a binary input gate (which is the equivalent of the Bulletproof's $\Phi$). Nevertheless, we do see clear differences, most noticeably in proof size. Our demonstration resulted in ZKP sizes of Bulletproofs at 480 bytes, zk-SNARKs at a deterministic 1,472 bytes, and post-quantum secure zk-STARKs averaging 18,342 bytes. In production, proofs with different customized constraints may be larger or smaller depending on the complexity of the constraints desired. Looking at the distributions of proof-generation time, we see Bulletproofs were by far the fastest, with a mean of 2.24 milliseconds. This is likely due to the limited complexity of Bulletproofs, which are optimized for zero-knowledge range proofs and offer limited flexibility otherwise. zk-SNARKs and zk-STARKs both had similar mean times, 79.73 and 72.86 milliseconds, respectively. zk-STARKs, however, had a substantially higher standard deviation, due to their non-deterministic nature. In terms of proof verification, the Bulletproof and zk-STARK implementations performed similarly, averaging around 1.1--1.2 milliseconds, with zk-SNARKs taking the longest at 9.70 milliseconds. We can thus conclude that Bulletproofs are the most efficient at the cost of arbitrary flexibility, zk-STARKs achieve post-quantum security at the expense of substantially bigger proofs, and zk-SNARKs represent a reasonable compromise despite requiring a trusted setup and having the slowest generation and verification times.

\begin{table}[!tbp]
    \centering
    \caption{ZKP Time and Size Comparisons}
    \label{tab:zkp-compare}
    \begin{tabular}{lccc}
        \toprule
        \textbf{Variant} & \textbf{ProofGen (ms)} & \textbf{ProofVerify (ms)} & \textbf{Package Size} \\
        \midrule
        Bulletproof & $2.24 \pm 0.91$ & $1.17 \pm 0.42$ & 924\,B \\
        zk-SNARK    & $79.73 \pm 22.41$ & $9.70 \pm 1.17$ & 1,981\,B \\
        zk-STARK    & $72.86 \pm 89.42$ & $1.14 \pm 0.26$ & 18,803\,B \\
        \bottomrule
    \end{tabular}
\end{table}

\subsection{Baseline Comparison}
\label{sec:baseline}

To quantify the cost of verifiability, we compare the full protocol against a baseline configuration using the same HE and threshold cryptography stack, but with ZKP generation and verification disabled. This baseline provides privacy and decentralization but no input verification. A malicious participant can encrypt any value without detection. Table~\ref{tab:baseline} reports the per-participant overhead introduced by each ZKP variant.

\begin{table}[!tbp]
    \centering
    \caption{Overhead of Verifiability: zk-DEAP vs.\ Unverified Baseline}
    \label{tab:baseline}
    \begin{tabular}{lcccc}
        \toprule
        \textbf{Metric} & \textbf{Baseline} & \textbf{BP} & \textbf{SNARK} & \textbf{STARK} \\
        \midrule
        Prove (ms)           & 0  & 2.24  & 79.73  & 72.86 \\
        Verify (ms)          & 0  & 1.17  & 9.70  & 1.14 \\
        Per-peer overhead (ms) & 0  & 3.40 & 89.43 & 74.01 \\
        Proof package size (bytes)   & 0    & 924   & 1,981 & 18,803 \\
        Round time, $n{=}100$ (ms) & 47.07 & 164.71 & 1,087.01 & 233.28 \\
        \midrule
        Verifiable           & \xmark & \cmark & \cmark & \cmark \\
        Anti-substitution    & \xmark & \cmark & \cmark & \cmark \\
        \bottomrule
    \end{tabular}
\end{table}

The baseline completes a round in 47.07\,ms per participant but provides no protection against value substitution or input poisoning. zk-DEAP with Bulletproofs adds 3.40\,ms per peer (prove + verify), providing cryptographic verification that eliminates input manipulation at its source. Unlike statistical defenses such as coordinate-wise trimmed mean~\cite{yin2018byzantine} or Krum~\cite{blanchard2017machine}, which degrade model accuracy and fail against sophisticated adaptive attacks~\cite{fang2020local}, cryptographic verification is exact, and no input violating $\Phi$ can enter the aggregate.
 
Each participant broadcasts a proof package to all $n-1$ peers, yielding $O(n^2)$ total messages. Table~\ref{tab:comm} reports per-participant and total communication for $n = 100$.
 
\begin{table}[!tbp]
\centering
\caption{Communication overhead per participant and total ($n = 100$, threshold $t = 50$).}
\label{tab:comm}
\begin{tabular}{lrrr}
\toprule
\textbf{Component} & \textbf{BP} & \textbf{SNARK} & \textbf{STARK} \\
\midrule
ElGamal ciphertext & 64\,B & 64\,B & 64\,B \\
Schnorr proof + binding data & 224\,B & 289\,B & 241\,B \\
ZKP & 480\,B & 1,472\,B & 18,342\,B \\
Ed25519 signature & 64\,B & 64\,B & 64\,B \\
Metadata & 92\,B & 92\,B & 92\,B \\
\midrule
\textbf{Proof package total} & \textbf{924\,B} & \textbf{1,981\,B} & \textbf{18,803\,B} \\
\textbf{Per-participant sent} & 91.5\,KB & 196.1\,KB & 1,861.5\,KB \\
\textbf{Network total} & 9.1\,MB & 19.6\,MB & 186.1\,MB \\
\midrule
Partial decryption pkg & $\sim$244\,B & $\sim$244\,B & $\sim$244\,B \\
Partial phase total & 1.21\,MB & 1.21\,MB & 1.21\,MB \\
\bottomrule
\end{tabular}
\end{table}
 
For comparison, SecAgg~\cite{bonawitz2017secagg} incurs $O(n^2)$ communication with per-participant cost of $O(n)$ masked seed shares (each 32\,B), totaling $\sim$8\,KB per participant for $n=100$, comparable to Bulletproof and SNARK variants. SecAgg+~\cite{bell2022secaggplus} reduces this to $O(n \log n)$ through sparse graph techniques, but by having a central server coordinating the masking graph. The sparser communication pattern is a direct consequence of relaxing the decentralization requirement. Our $O(n^2)$ broadcast cost is the price of full decentralization. Each participant must independently verify every other participant's proof, forcing a full-mesh exchange. The core zk-DEAP protocol sits at a polar end of the decentralization-communication trade-off. That said, coordinator-assisted deployments can still compose with the protocol and route proof packages through a single relay (reducing per-participant communication to $O(1)$). Furthermore, multi-tier hierarchical deployments can run full-mesh zk-DEAP within local groups of modest size and pass aggregates upward, recovering near-linear total communication while preserving decentralization within each tier.

\subsection{Network Scaling}

\begin{figure}[t]
    \centering
    \includegraphics[width=0.4\textwidth]{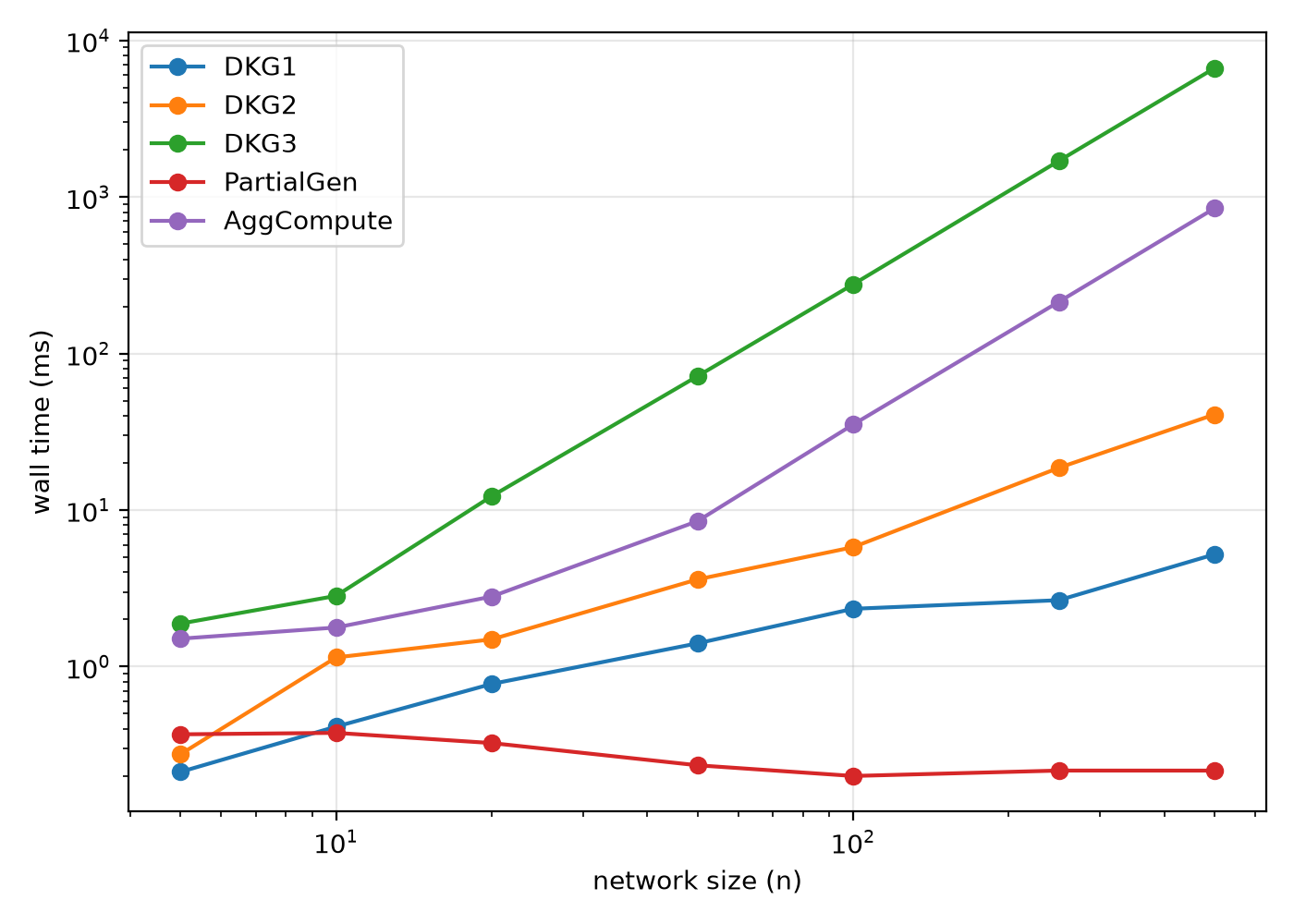}
    \caption{Network Size vs.~Wall Time Comparison of Network-Dependent Functions.}
    \label{fig:n-vs-time}
\end{figure}

We observe the performance of the network-dependent functions in Figure~\ref{fig:n-vs-time}.

The DKG setup phase exhibits superlinear growth with network size, while the core cryptographic operations (proof generation, verification, and partial decryption) scale efficiently. FROST phase 2 (encompassing functions DKG2 and DKG3), in particular, with an $O(n^2t+nt^2)$ cost, where $n$ is the number of participants and $t$ is the threshold, starts to become a noticeable computational bottleneck in large networks with large thresholds, as each participant needs to send pairwise messages to each other participant. Fortunately, the FROST DKG setup only runs once at the start of the protocol and is not necessary once keys are established. Looking at the other network-dependent functions, we see that, although it is currently dwarfed by the DKG, aggregate computation also grows with network size. This makes sense given the necessity of the BSGS algorithm to extract the aggregate from the ElGamal ciphertext. For reasonable message spaces, it is efficient and practical, only presenting a roadblock in large networks in the thousands or large aggregates in the millions, where a windowed-BSGS can be used~\cite{menezes1996handbook}.

Per-participant cost is linear, and the network total is quadratic, so at $n=500$, the Bulletproof variant sends 0.46\,MB per participant and 0.23\,GB for the whole network, while the zk-STARK variant sends 9.4\,MB per participant and over 4\,GB for the whole network. Blockchain and large-scale voting, at over $10^4$ participants, may require the 
hierarchical scheme described in Section~\ref{sec:baseline}, while most FL, smart-grids, and IoT deployments below a few hundred participants can use the zk-DEAP protocol as described in Section~\ref{sec:bindmechref}.

\subsection{Federated Learning Deployment Context}
\label{sec:fl-context}
To demonstrate practical deployability, Table~\ref{tab:fl-context} maps zk-DEAP's measured performance to FL parameters. This helps contextualize how the protocol performs on a real, production task that requires privacy, verifiability, and decentralization. Cryptographic overhead is dominated by peer verification, which is parallel across cores. Our single-thread evaluation represents a conservative floor. DKG cost is amortized across FL deployments lasting hours or days~\cite{kairouz2021advances}.

\begin{table}[!tbp]
    \centering
    \caption{zk-DEAP Performance Mapped to FL Deployment (Single thread, $n{=}100$, Bulletproof Variant)}
    \label{tab:fl-context}
    \begin{tabular}{llr}
        \toprule
        \textbf{FL Phase} & \textbf{zk-DEAP Component} & \textbf{Time} \\
        \midrule
        Cluster formation   & FROST DKG (one-time)       & $\sim$60\,s \\
        Input submission    & Prove            & $\sim$2.2\,ms \\
        Peer verification   & Verify $\times$ 99 peers   & $\sim$115.4\,ms \\
        Secure aggregation  & Homomorphic multiply       & $<$1\,ms \\
        Result recovery     & PDec + Combine + BSGS      & 35.4\,ms \\
        \midrule
        \textbf{Total crypto/round} &           & $\sim$\textbf{153.0\,ms} \\
        Typical FL round~\cite{bonawitz2019towards} & Local train + comms & 10--60\,s \\
        \textbf{Relative overhead} &             & $\sim$\textbf{0.25\%} (60 s) \\
        \bottomrule
    \end{tabular}
\end{table}

The current implementation operates on scalar inputs, which directly serves the primary use cases of sensor aggregation, vote tallies, and scalar model metrics. For high-dimensional FL gradients, zk-DEAP extends naturally. Each gradient coordinate could be encrypted and proven independently, with proofs batched for verification. At $d=1000$ dimensions, the Bulletproof variant's naive per-coordinate extrapolation yields ${\sim}3.4$\,s total. This is a pessimistic upper bound, as native batched Bulletproofs over shared generators amortize the inner-product argument and run substantially faster, as do the specialized vector-level schemes (Section~\ref{sec:related}). Approaches such as random linear projections~\cite{risefl2024}, Pedersen vector commitments with batched Bulletproofs~\cite{acorn2023}, or lattice-based ZKPs~\cite{lzksa2025} compose with our binding mechanism and represent promising optimizations in certain FL deployments. Crucially, our core contribution is generic and dimension-independent. The binding machinery applies per-coordinate without modification. For networks exceeding several hundred participants, the FROST DKG becomes the bottleneck due to $O(n^2 t + nt^2)$ pairwise communication. Arbitrarily large-scale deployments can employ hierarchical architectures where local groups run zk-DEAP independently and pass aggregates to the next tier, trading decentralization for scalability. Cross-institutional FL deployments, however, typically involve 10-200 organizations, well within efficient operating range.

The protocol assumes participants with sufficient computational resources for elliptic curve operations and ZKP generation. Low-powered sensors with constrained CPUs would require gateway-mediated participation. The full-mesh broadcast topology is most efficient for networks under several hundred participants. Sparse network architectures would benefit from relay-assisted message distribution.

\subsection{Evaluation Scope}
\label{sec:eval-scope}
Our per-operation evaluation matches NDSS-style cryptographic protocol papers~\cite{threshold-ecdsa-ndss2024,cirrus-ndss2026,cwpsu-ndss2026} and reflects three deliberate choices. First, we do not reimplement RoFL, ACORN, or LZKSA on identical hardware, as they target the centralized single-server setting with materially different security properties, so matched-hardware numbers would compare systems solving different problems. Second, we measure the cryptographic layer directly rather than embedding zk-DEAP in a specific FL framework, since the protocol is general-purpose (voting, smart grids, IoT, FL) and framework-specific numbers would narrow that scope. As such, we report per-scalar rather than per-vector measurements because anti-substitution binding is dimension-independent. Vector batching composes with, rather than alters, our mechanism. Third, single-threaded numbers are a conservative floor, and operations parallelize linearly. Our goals were to emphasize clarity and generality, so that anyone wishing to adopt zk-DEAP (in any domain) can assess the performance implications.

\section{Related Work}
\label{sec:related}

\subsubsection{Secure Aggregation} Table~\ref{tab:related} compares zk-DEAP against existing protocols on several properties, including privacy, verifiability, and decentralization. 

SecAgg~\cite{bonawitz2017secagg} and SecAgg+~\cite{bell2022secaggplus} established practical privacy-preserving aggregation for FL through pairwise masking, while Flamingo~\cite{flamingo2023} extended this to multi-round settings with efficient dropout resilience. None verify input constraints. VeriFL~\cite{verifl2021} introduced communication-efficient verification of server aggregation integrity using homomorphic hashing, but does not verify that client inputs satisfy a predicate $\Phi$, a fundamentally different problem.

EIFFeL~\cite{eiffel2022} was the first framework to enforce arbitrary integrity constraints on secret-shared FL updates using SNIP proofs. RoFL~\cite{rofl2023} used Bulletproofs to enforce norm bounds on encrypted gradients, and ACORN~\cite{acorn2023} improved on this with Pedersen vector commitments and inner product arguments for input validation. ELSA~\cite{elsa2023} achieved up to $305\times$ speedup over RoFL through distributed trust across two servers with MPC-based norm checks. RiseFL~\cite{risefl2024} further improved efficiency using hybrid commitments with random linear projections, achieving up to $28\times$ speedup over ACORN. Most recently, LZKSA~\cite{lzksa2025} introduced lattice-based ZKPs for post-quantum secure input verification, verifying $L_\infty$, $L_2$ norms and cosine similarity constraints. All of these achieve privacy and verifiability but require trusted servers. None are fully decentralized.

Specialized vector-level aggregation dominates zk-DEAP for high-dimensional FL. For $L_\infty$ on CIFAR10-scale inputs, RoFL~\cite{rofl2023} reports $\sim$94\,s/29.9\,s prove/verify, ACORN~\cite{acorn2023} reports 78.7\,s/33.9\,s, and LZKSA~\cite{lzksa2025} achieves 0.02\,s/0.0062\,s via lattice packing. zk-DEAP's Bulletproof variant at 2.24\,ms/1.17\,ms per scalar buys a different property with this overhead, elimination of the trusted coordinator. For zero-trust federated learning across mutually distrusting servers, voting, smart-grids, IoT/OT environments, or blockchain protocols, zk-DEAP is the first scheme to provide privacy, verifiability, and anti-substitution binding simultaneously. Server-optimized schemes remain optimal when centralization is acceptable.

The majority of the Decentralized Federated Learning (DFL) scholarship~\cite{lalitha2019peer, warnat2021swarm, beltran2024fedstellar} has examined properties such as convergence and scheduling, rather than the mechanics of how neighbors securely communicate. Existing approaches often employ plaintext updates to neighbors, introducing inversion attacks~\cite{zhu2019deep}, or use pairwise masking, which helps with privacy but not verifiability. This motivates the need for a primitive such as zk-DEAP, which can simultaneously address the aforementioned requirements of decentralized security.

Among protocols targeting decentralization, PrivDA~\cite{privda2021} used Paillier HE with Ethereum but relied on a trusted aggregator and lacked ZKP verification. PPVDA~\cite{wang2024ppvda} achieved privacy and decentralization through homomorphic MACs but without constraint verification. DC-nets FL~\cite{deoliveira2024dcnets} and GVSA~\cite{zhou2025gvsa} used secret sharing for decentralized FL but provided neither HE-level privacy nor ZKP verification. TrustDFL~\cite{wang2024trustdfl} employed Groth16 ZKPs with blockchain coordination, achieving verifiability without input privacy. Zhang et al.~\cite{ZHANG2022103118} combined HE with ZKPs for smart grid aggregation but required a trusted aggregator. zk-DEAP is the first protocol to achieve all three properties simultaneously, with a binding mechanism that prevents substitution attacks, to which any protocol combining independent HE and ZKP components is otherwise vulnerable.

\subsubsection{Theoretical Foundations} Bendlin and Damg{\aa}rd~\cite{BendlinD10} introduced threshold decryption with ZKPs for lattice-based cryptosystems, focusing on general threshold PKE rather than aggregation. Boneh et al.~\cite{BonehGGJKRS18} presented a framework for building threshold cryptosystems from threshold FHE. Zooming out, multi-key HE~\cite{lopez2012fly} allows for independently keyed ciphertexts, and multiparty or threshold HE~\cite{mouchet2021multiparty} allows for a distributed key, both on arbitrary circuits, but neither provides any verifiability. Xie et al.~\cite{XieHuangQiu25} proposed verifiable threshold multi-party FHE using commitment and NIZK techniques. These works establish theoretical foundations, while zk-DEAP addresses the distinct engineering constraints of practical aggregation with concrete efficiency requirements. In threshold cryptography specifically, FROST~\cite{KomloG20,RFC9591} has seen adoption primarily for blockchain signatures~\cite{Komatireddy2025} rather than aggregation. zk-DEAP is the first to integrate FROST with HE and ZKPs for this purpose.

\subsubsection{Cross-domain ZKP linking} The technical core of our contribution is the cross-prime case left open by the existing literature on linking ciphertexts or commitments to ZKPs over committed values (LegoSNARK~\cite{campanelli2019legosnark} for Pedersen-committed gadgets, SAVER~\cite{lee2019saver} for ElGamal-compatible Groth16, Eclipse~\cite{aranha2022eclipse} and Tangram~\cite{cryptoeprint:2025/540} for PLONK-style systems, and Zether's $\Sigma$-Bullets~\cite{bunz2020zether} and PGC's twisted ElGamal~\cite{chen2020pgc} for Bulletproof linkage). All rely on the commitment group and proof system sharing an algebraic structure (same field, or fields related by a known pairing). zk-DEAP's Bulletproof variant falls in this pattern and uses a correspondingly lightweight randomness-shift fix. The Ristretto255 scalar field, however, is unrelated to Halo2's or Winterfell's by any embedding, and embedding foreign-field arithmetic in-circuit is highly inefficient. Other attempts at non-interactive ZKP composition~\cite{chase2016efficient, agrawal2018nizk, backes2019cross, tu2026lightweight} address algebraic/non-algebraic statement linking more generally, but are either interactive, SNARK-specific with trusted setup, or built on dedicated inflexible proof systems. Our blinded pre-challenge MiMC commitment with explicit carry targets this cross-prime gap.

\subsubsection{Complementary Defenses}
Differential privacy~\cite{dwork2006calibrating,stevens2022usenix} provides output privacy orthogonal to zk-DEAP's input privacy and integrity guarantees. The two compose naturally. Statistical Byzantine defenses such as Krum~\cite{blanchard2017machine} and coordinate-wise trimmed mean~\cite{yin2018byzantine} filter malicious updates heuristically but degrade model accuracy and fail against adaptive attacks~\cite{fang2020local}. zk-DEAP provides exact verification regardless of adversarial strategy.

\section{Conclusion}
\label{sec:conclusion}

We presented zk-DEAP, the first secure aggregation protocol achieving simultaneous privacy, verifiability, decentralization, and anti-substitution binding.

The protocol is designed for general-purpose additive aggregation, including weighted sums and linear combinations, and applies to federated learning, IoT/OT telemetry, smart grids, electronic voting, blockchains, and any domain requiring decentralized, verifiable computation of private sums. The current design is limited to additive operations over a bounded message space (constrained by BSGS extraction), and communication scales as $O(n^2)$ due to coordinator-free broadcast. Future work includes parallelized verification, batched vector proofs for high-dimensional FL gradients, hierarchical architectures for large-scale deployments, alternative HE primitives to circumvent the BSGS bottleneck, and post-quantum extensions beyond the STARK variant. Our implementation is open-source at \href{https://doi.org/10.6084/m9.figshare.33411562} {https://doi.org/10.6084/m9.figshare.33411562}.

\section*{Acknowledgments}
We would like to thank the Colgate Department of Computer Science for their support throughout the project. We would also like to thank David Perkins and Kelly Isham for their feedback and discussion.

\textbf{Generative AI Usage.} To facilitate testing, \href{https://claude.ai}{claude.ai} was used in both the fixture server and test harness code. Specifically, after manually implementing and verifying one test case, Claude was prompted to replicate the case under explicitly chosen parameters to draft new tests. The resulting code was manually verified before being used. Claude was similarly employed to draft the script that analyzed the results and produced visuals. Once again, it was prompted with specific goals and manually verified before being used. To promote transparency, all files in the artifact containing AI-generated code begin with a disclaimer on the first line.

\bibliographystyle{IEEEtran}
\bibliography{main}

\appendices

\section{Formal Security Proof of zk-SNARK and zk-STARK Binding}
\label{app:stark-snark-proof}

\subsection{Theorem Statement}
 
Under the discrete logarithm assumption on Ristretto255 and the following MiMC key-recovery under known plaintext assumption, no PPT adversary can produce a valid proof package where the ElGamal ciphertext encrypts $s_i$ with $\Phi(s_i) \neq 0$, except with negligible probability.

\textbf{Assumption (MiMC$_{63}$ key-recovery under known plaintext, single query).} For any fixed $(s, \epsilon) \in \mathbb{F}_p^2$ and any PPT adversary $\mathcal{A}$,
\begin{multline}
\Pr\![\,\mathcal{A}(s, \epsilon, v, V) \to b' \text{ s.t.\ } \mathrm{MiMC}_{63}(v; s, \epsilon, b') = \\
V\;\Big|\; b, v \overset{\$}{\leftarrow} \mathbb{F}_p,\; V \!:=\! \mathrm{MiMC}_{63}(v; s, \epsilon, b)\,] \le \mathsf{negl}(\lambda).
\end{multline}
That is, given a single plaintext-ciphertext pair $(v, V)$ under an unknown blinding key $b$, no PPT adversary can recover any equivalent key $b'$ producing the same image. Note that $v$ is an input to $\mathcal{A}$, not an output, so the adversary cannot invert MiMC under a chosen key, because the protocol fixes $v$ via the Schnorr-linear constraint after the Fiat-Shamir challenge is committed. For our 63-round construction with $x^5$ S-box over $|\mathbb{F}_p| \approx 2^{128}$, the best known attack requires $\Omega(2^{128})$ work, as analyzed in the assumption-characterization subsection below.
 
\subsection{Notation}
 
\begin{itemize}
\item \textbf{Group}: Ristretto255 with generator $g$, order $q \approx 2^{252}$
\item \textbf{DKG}: $Y = g^x$ (public key), $x$ distributed via FROST
\item \textbf{Field}: $\mathbb{F}_p$ where $p = 2^{128} - 45 \cdot 2^{40} + 1$ (Winterfell prime, for Halo2, $\mathbb{F}_r$ with $r \approx 2^{254}$)
\item \textbf{Ciphertext}: $(c_1^{(i)}, c_2^{(i)}) =(g^{r_i}, g^{s_i}  Y^{r_i})$
\item \textbf{Schnorr nonce}: $v \overset{\$}{\leftarrow} \mathbb{Z}_q$, reduced as $v_f = v \bmod p$
\item \textbf{Blinding factor}: $b \overset{\$}{\leftarrow} \mathbb{F}_p$
\item \textbf{MiMC}: $\mathrm{MiMC}_{63}(x;\; \mathbf{k})$ denotes 63 rounds of $F(x) = (x + k_i)^5$ where $k_i = s_\text{ZKP} + \epsilon_i + b + \mathit{rc}_i$
\item \textbf{Commitment}: $V_\text{commit} = \mathrm{MiMC}_{63}(v_f;\; s_\text{ZKP}, \epsilon_i, b)$, appended to transcript before challenge
\item \textbf{Public inputs}: $V_\text{commit}$, $\text{resp}_f$, $e_f$, $\epsilon_i$
\item \textbf{Private witnesses}: $s_\text{ZKP}$, $v_f$, $b$, $\omega \in [0, S_\text{max}]$
\item \textbf{Valid set}: $\mathcal{S}_\Phi = \{s \in \mathbb{F}_p : \Phi(s) = 0\}$ ($\Phi$ may be arbitrary)
\end{itemize}
 
\subsection{MiMC Permutation Property}
\label{app:mimc_perm}
 
For fixed round keys $\mathbf{k} = (k_0, \ldots, k_{62})$, define $\mathrm{MiMC}_\mathbf{k} : \mathbb{F}_p \to \mathbb{F}_p$ as the composition of 63 round functions $F_i(x) = (x + k_i)^5$. Each
$F_i$ is a bijection on $\mathbb{F}_p$ because:
\begin{enumerate}
\item Adding a constant is a bijection.
\item $x \mapsto x^5$ is a bijection when $\gcd(5, p - 1) = 1$. For $p = 2^{128} - 45 \cdot 2^{40} + 1$: $p - 1 \equiv 0 \pmod{2}$ and $p - 1 \bmod 5$: since $2^4 \equiv 1 \pmod{5}$, we have $2^{128} = (2^4)^{32} \equiv 1 \pmod{5}$, so $p - 1 \equiv 1 - 45 \cdot 2^{40} \equiv 1 - 0 \equiv 1 \pmod{5}$ (since $45 \equiv 0 \pmod{5}$). Thus $\gcd(5, p-1) = \gcd(5, 1) = 1$.
\end{enumerate}
The composition of bijections is a bijection. Therefore, $\mathrm{MiMC}_\mathbf{k}$ is a permutation: distinct inputs produce distinct outputs under the same key schedule.
 
\subsection{Security Game}
 
The adversary $\mathcal{A}(\lambda)$ receives $(g, Y)$ where $Y = g^x$ for unknown $x$, and outputs a proof package containing:
\begin{itemize}
\item ElGamal ciphertext $(c_1, c_2)$
\item Schnorr proof $\pi_\text{schnorr}$ with response $s_\text{resp}$
\item ZKP $\pi_\text{ZKP}$ with public inputs $(V_\text{commit}, \text{resp}_f, e_f, \epsilon_i)$
\end{itemize}
such that $\pi_\text{schnorr}$ and $\pi_\text{ZKP}$ both verify. $\mathcal{A}$ wins if $(c_1, c_2)$ encrypts $s$ with $\Phi(s) \neq 0$.
 
\subsection{Proof}
 
\textbf{Step 1: Witness Extraction.}
By the special soundness of $\Sigma$-protocols (two accepting transcripts with challenges $e \neq e'$), extract from
$\pi_\text{schnorr}$:
\[
  s_\text{schnorr},\; r \quad\text{such that}\quad c_1 = g^r,\; c_2 = g^{s_\text{schnorr}}  Y^r
\]
By the knowledge soundness of the ZKP system, extract from $\pi_\text{ZKP}$:
\[
  s_\text{ZKP},\; v_f,\; b,\; \omega \quad\text{satisfying Equations (4-6)}
\]

Extraction in the non-interactive setting follows from the forking lemma~\cite{pointcheval2000security} applied to the Fiat-Shamir transcript, with concrete security loss $\epsilon_\text{fork} \leq q_H / 2^{256}$ for $q_H$ random oracle queries, which is negligible.
 
\textbf{Step 2: Schnorr Response Consistency.}
The Fiat-Shamir challenge $e$ is derived from a transcript including $V_\text{commit}$. The Schnorr response is $s_\text{resp} = v + e  s_\text{schnorr}$ (in $\mathbb{Z}_q$), giving $\text{resp}_f = s_\text{resp} \bmod p$.
 
From Equation (5): $\text{resp}_f = v_f + e_f  s_\text{ZKP} - \omega  \tilde{q}$ (in $\mathbb{F}_p$ for STARK, in $\mathbb{F}_r$ for SNARK), with $\omega \in [0, S_\text{max}]$ enforced within the circuit/AIR in both variants.
 
Therefore: $v_f = \text{resp}_f - e_f  s_\text{ZKP} + \omega  \tilde{q}$.
 
If $s_\text{ZKP} = s_\text{schnorr}$, then $v_f = v \bmod p$ (the true Schnorr nonce), and Equation (6) is satisfied with the original $b$.
 
If $s_\text{ZKP} \neq s_\text{schnorr}$, define $\delta = s_\text{schnorr} - s_\text{ZKP} \neq 0$ and:
\[
  v_f' = v_f + e_f  \delta + (\omega' - \omega)  \tilde{q}
\]
where $\omega'$ is the adversary's chosen carry, range-checked to $[0, S_\text{max}]$.

The adversary committed $V_\text{commit} = \mathrm{MiMC}_{63}(v_f;\; s_\text{ZKP}, \epsilon_i, b)$ before the challenge $e$ (and hence $e_f$) was derived.
 
\textbf{Step 3: Case Analysis.}
Recall from Step 2 that $v_f' = v_f + e_f  \delta + (\omega' - \omega)  \tilde{q} \pmod p$ with $\delta = s_\text{schnorr} - s_\text{ZKP} \neq 0$. The adversary must satisfy $\mathrm{MiMC}_{63}(v_f';\; s_\text{ZKP}, \epsilon_i, b') = V_\text{commit}$ for some $(\omega', b')$ with $\omega' \in [0, S_\text{max}]$. We can split on whether $v_f' = v_f$ and whether $b' = b$.
 
\textit{Case (i): $v_f' = v_f$ (any $b'$).} This requires $e_f  \delta + (\omega' - \omega_\mathrm{hon})  \tilde{q} \equiv 0 \pmod p$, where $\omega_\mathrm{hon} = \lfloor (v + e  s_\text{schnorr})/q \rfloor$ is the carry implied by the extracted Schnorr witness (a specific value in $[0, S_\text{max}]$, not a free variable). Setting $v_f' = v_f$ is therefore one linear equation in $e_f$ per choice of $\omega' \in [0, S_\text{max}]$, with nonzero coefficient $\delta \in \mathbb{F}_p$ (since $|\delta| \le S_\text{max} \ll p$ implies $\delta$ is a unit in $\mathbb{F}_p$). Under the ROM each equation holds with probability $1/p$ over uniform $e_f$. Union bound over the $S_\text{max} + 1$ admissible values of $\omega'$ gives $(S_\text{max} + 1)/p$ per RO query; over $q_H$ Fiat-Shamir queries, $q_H  (S_\text{max} + 1)/p$. $\square$
 
\textit{Case (ii): $v_f' \neq v_f$ and $b' = b$.} The round keys are identical, so $\mathrm{MiMC}_{63}{(\cdot;\, s_\text{ZKP}, \epsilon_i, b)}$ is a permutation on $\mathbb{F}_p$ (Section~\ref{app:mimc_perm}). A permutation maps distinct inputs to distinct outputs, so $\mathrm{MiMC}_{63}(v_f';\, s_\text{ZKP}, \epsilon_i, b) \neq \mathrm{MiMC}_{63}(v_f;\, s_\text{ZKP}, \epsilon_i, b) = V_\text{commit}$. Probability $0$. $\square$
 
\textit{Case (iii): $v_f' \neq v_f$ and $b' \neq b$.} The adversary outputs a key $b' \neq b$ with $\mathrm{MiMC}_{63}(v_f';\, s_\text{ZKP}, \epsilon_i, b') = V_\text{commit}$, where $V_\text{commit}$ was fixed before $e_f$ (and hence $v_f'$) was known. We construct a reduction $\mathcal{B}_2$ to the MiMC key-recovery assumption: $\mathcal{B}_2$ receives a challenge $(s_\text{ZKP}, \epsilon_i, v', V)$, embeds $v_f := v'$ into the simulated transcript and $V_\text{commit} := V$, runs the binding adversary, and outputs the extracted $b'$. Since $\mathcal{B}_2$ wins exactly when the adversary's $b'$ satisfies the MiMC equation under the chosen $v' = v_f'$, this case is bounded by $\mathrm{Adv}_\mathrm{MiMC}^\mathrm{KR\text{-}KP}(\mathcal{B}_2)$. $\square$

In all three cases, $s_\text{ZKP} \neq s_\text{schnorr}$ leads to negligible advantage. Therefore $s_\text{ZKP} = s_\text{schnorr} = s$, and $\Phi(s) = 0$ by Equation (4).
 
\textbf{Step 4: Ciphertext Consistency.} As in the Bulletproof variant, Equations (1) and (2) with shared $r_\text{resp}$ ensure $c_1 = g^r$ and $c_2 = g^s  Y^r$ for the same $r$, preventing randomness substitution (Attack 2). $\square$

\textbf{Concrete Bound.} Combining all reduction steps, the binding advantage of any PPT adversary $\mathcal{A}$ making at most $q_H$ random oracle queries satisfies:
\begin{multline}
\mathrm{Adv}_{\Pi}^{\mathrm{bind}}(\mathcal{A}) \;\leq\; \mathrm{Adv}_{\mathrm{Rist255}}^{\mathrm{dlog}}(\mathcal{B}_1) + \mathrm{Adv}_{\mathrm{MiMC}}^{\mathrm{KR\text{-}KP}}(\mathcal{B}_2) \\
+\; \frac{q_H  (S_{\max} + 1)}{p} \;+\; \frac{q_H}{2^{256}}
\end{multline}
where $\mathcal{B}_1$ is the DLP solver from the Schnorr extractor (Step 1), $\mathcal{B}_2$ is the MiMC key-recovery adversary (Step 3, Case (iii)), the $q_H  (S_{\max} + 1)/p$ term bounds the Case~(i) carry-collision probability over $q_H$ Fiat-Shamir queries, and $q_H / 2^{256}$ is the Fiat-Shamir forking-lemma loss (Step 1). For the implemented binary constraint ($S_{\max}=1$) with $|\mathbb{F}_p| \approx 2^{128}$ and $q_H \le 2^{64}$, the carry term is bounded by $2^{-63}$; for an 8-bit constraint ($S_{\max}\le 2^8$) it is bounded by $2^{-56}$. The bound thus dominates the security parameter for large input domains and is the binding-significant term. $\square$
 
\subsection{Zero-Knowledge}
 
The private witnesses $s_\text{ZKP}$, $v_f$, $b$, $\omega$ have no boundary assertions in the STARK trace (resp.\ are private witnesses in the Halo2 circuit). The ZKP system's zero-knowledge property ensures the verifier learns nothing beyond the public inputs. The blinding factor $b$ provides an additional layer. Even given $V_\text{commit}$, $\text{resp}_f$, and $e_f$, the verifier cannot determine $s_\text{ZKP}$ without recovering $b$, which is the MiMC key-recovery game and requires $2^{128}$ work.
 
\subsection{Generalization}
 
The binding mechanism is constraint-independent up to the carry width. Any $\Phi(s_\text{ZKP}) = 0$ whose witness domain $\mathcal{S}_\Phi$ satisfies $S_\text{max} \ll p$ binds to the encrypted value $s_i$, because the embedded Schnorr constraint (Equation (5)) and blinded hash commitment (Equation (6)) force $s_\text{ZKP} = s_i$ regardless of what $\Phi$ is, provided $\omega \in [0, S_\text{max}]$ is range-checked. This covers gradient norm bounds, cosine similarity constraints, and any other practically relevant predicate expressible in the circuit/AIR.
 
\subsection{Modular Reduction via Explicit Carry Value}

The Schnorr response $s_\text{resp} = (v + e  s) \bmod q$ is computed in $\mathbb{Z}_q$, while the STARK operates in $\mathbb{F}_p$. Writing $\phi: \mathbb{Z}_q \to \mathbb{F}_p$ for canonical reduction, we have $\phi(a + b \bmod q) \neq \phi(a) + \phi(b)$ in general: they differ by $\phi(q) = \tilde{q}$ precisely when $a + b \geq q$ in $\mathbb{Z}$.

To handle this, the prover includes a carry value $\omega = \lfloor (v + e  s_\text{ZKP})/q \rfloor$ as a private witness. Since $v, e \in [0, q)$ and $s_\text{ZKP} \in \mathcal{S}_\Phi \subseteq [0, S_\text{max}]$, we have $\omega \in [0, S_\text{max}]$. The AIR/circuit enforces Equation (5) and range-checks $\omega \in [0, S_\text{max}]$ via an in-proof constraint on the private witness. For our implemented binary constraint $\mathcal{S}_\Phi = \{0,1\}$, this reduces to $\omega \in \{0,1\}$, a single bit. For a $k$-bit range constraint, $\omega$ is $k$ bits wide.

\textbf{Soundness.} Suppose the adversary produces a valid Schnorr proof for $s_\text{schnorr}$ but wants the STARK to prove $s_\text{ZKP} \neq s_\text{schnorr}$. The constraint then requires
\[
\omega = \frac{v_f + e_f  s_\text{ZKP} - \text{resp}_f}{\tilde{q}} \pmod p,
\]
which, for Fiat-Shamir challenge $e$ modeled as a random oracle output, is a pseudorandom element of $\mathbb{F}_p$ (since $\text{resp}_f$ depends linearly on $e_f$ with coefficient $\delta \neq 0$). The probability the required $\omega$ lands in $[0, S_\text{max}]$ is at most $(S_\text{max}{+}1)/p$. For $p \approx 2^{128}$ this is negligible for any practically relevant constraint. Our binary case gives $2/p \approx 2^{-127}$, an 8-bit range gives $2^{-120}$, and even a 64-bit range gives $2^{-64}$.

Thus the modular reduction is handled with full binding security with the same explicit-$\omega$ construction in both the SNARK and STARK variants. The binding applies to arbitrary $\Phi$ whose witness domain satisfies $S_\text{max} \ll p$, covering all practically relevant predicates including $k$-bit ranges, gradient-norm bounds, and cosine-similarity checks.

\subsection{Characterization of the MiMC Key-Recovery Assumption}

The assumption instantiates a single-query key-recovery game under known plaintext. The adversary knows input $v_f'$ (determined after commitment by $e_f$), output $V_\text{commit}$ (committed before~$e_f$), and values $s_\text{ZKP},\epsilon_i$, and must find $b'$ such that $\mathrm{MiMC}_{K'}(v_f')=V_\text{commit}$ where $K'=s_\text{ZKP}+\epsilon_i+b'$. This is different from standard preimage resistance, which fixes the key and varies the input. Our game fixes the input and varies the key. These are complementary formulations of one-wayness for keyed permutations.

For fixed $v_f'$, the output is a univariate polynomial in $K'$ of formal degree $5^{63}\approx 2^{146}$, reducing to at most $p-1\approx 2^{128}$ by Fermat's little theorem. Berlekamp-Cantor-Zassenhaus factorization~\cite{von2013modern} costs $O(p\cdot\mathrm{polylog}(p))$, exceeding brute-force key search at $O(p)\approx O(2^{128})$ MiMC evaluations, so generic enumeration is the dominant strategy. The principal algebraic attacks against $x^5$-MiMC (interpolation requiring $\geq 5^{\lceil r/2\rceil}$ chosen plaintexts~\cite{cryptoeprint:2016/492}, higher-order differentials~\cite{eichlseder2020algebraic}, and GCD-based key recovery~\cite{albrecht2020algebraic}) all require multiple plaintext-ciphertext pairs under a fixed unknown key, a strictly more permissive setting than our single-query game. The $O(2^{128})$ generic bound therefore applies, and our 63 rounds exceed the recommended minimum of $\lceil 128/\log_2 5\rceil=56$~\cite{cryptoeprint:2016/492}.

\section{Artifact Appendix}
\label{app:artifact}

This appendix describes the artifact associated with this paper according to the NDSS Artifact Evaluation process. 

\subsection{Description \& Requirements}

The artifact includes our Rust implementation of zk-DEAP with all three proof-system variants,
five-participant demonstrations, 60 attack test cases, the benchmark harness, and a
script that regenerates the tables and Figure~\ref{fig:n-vs-time} of Section~\ref{sec:testing}.

\subsubsection{How to access}
The artifact is published on Figshare: \url{https://doi.org/10.6084/m9.figshare.33411562}

\subsubsection{Hardware dependencies}
Any x86-64 or arm64 machine running Linux or macOS. About 1.5\,GB of disk for a native
build or 4\,GB for the Docker image. Under 100\,MB of memory at the default participant
counts; the full Figure~\ref{fig:n-vs-time} sweep needs more, as the fixture server holds every
pre-computed configuration in memory at once.

\subsubsection{Software dependencies}
Rust 1.88 or newer, \texttt{git}, and Python~3 with matplotlib. On
Linux, \texttt{build-essential} and \texttt{pkg-config} with
\texttt{libssl-dev} for \texttt{openssl-sys}. On macOS, the Xcode command line tools.

\subsubsection{Benchmarks}
No dataset, model, or external workload is required. The fixture server generates the
inputs deterministically, and every reported value comes from a fresh run.

\subsection{Artifact Installation \& Configuration}

Using Docker:

\begin{samepage}
\begin{verbatim}
docker build -t zk-deap .
docker run --rm zk-deap
\end{verbatim}
\end{samepage}

To keep the CSV tables and Figure~\ref{fig:n-vs-time}, run without \verb|--rm| and copy them out:

\begin{samepage}
\begin{verbatim}
docker run --name zk-deap-run zk-deap
docker cp zk-deap-run:/zk-deap/results \
    ./results
docker rm zk-deap-run
\end{verbatim}
\end{samepage}

Alternatively, a native build:

\begin{samepage}
\begin{verbatim}
cargo build --release --locked
python3 -m venv .venv
./.venv/bin/pip install matplotlib
./run_all.sh
\end{verbatim}
\end{samepage}

Either set of commands will run all three experiments. More detailed information is provided in the artifact README.

\subsection{Major Claims}

\begin{itemize}
  \item (C1): All three variants complete the protocol end to end, and every participant
  independently recovers the same correct aggregate (E1).

  \item (C2): The binding mechanism prevents the substitution attacks from Section~\ref{sec:attttack},
  and the implementation rejects the further attacks from Section~\ref{sec:shared-attacks} (E2).

  \item (C3): The Bulletproof variant has the smallest proofs and is the fastest proving. The zk-STARK variant has much
  larger proofs, and the zk-SNARK variant is in between on proof size (E3). Reported in Tables~\ref{tab:zkp-compare}~and~\ref{tab:comm}.
  
  \item (C4): Verification cost is dominated by checking peers' proofs, and the per-round
  cryptographic overhead is a small fraction of a federated-learning round (E3). Reported in
  Tables~\ref{tab:baseline}~and~\ref{tab:fl-context}.

  \item (C5): Cost of the network-dependent operations grows with participant count as
  Figure~\ref{fig:n-vs-time} reports, dominated by the distributed key generation (E3).
\end{itemize}

\subsection{Evaluation}

Experiments are run with \texttt{run\_all.sh} (directly or as the Docker default command). Expect $<$10 minutes for the run.

\subsubsection{Experiment (E1): The protocol functions}
Each proof variant runs five participants with inputs $1,0,1,1,0$, so every
participant must print \texttt{sum=3/5}. Any other sum refutes C1.

\subsubsection{Experiment (E2): Attacks are rejected}
The four test files hold 13, 15, 16, and 16 attack cases, 60 in all, and each
should print \texttt{RESULTS: $n$ passed, 0 failed}. A failure means an
attack succeeded and refutes C2.

\subsubsection{Experiment (E3): Per-operation cost}
A local fixture server supplies deterministic key material, and the harness times each
protocol operation against it, thirty runs per configuration. It prints Tables~\ref{tab:zkp-compare}--\ref{tab:fl-context}, also
writing them and Figure~\ref{fig:n-vs-time} to \texttt{results/}. Absolute times scale with the hardware, so what should hold are the
orderings claimed in C3 to C5: ZKP sizes Bulletproof $<$ zk-SNARK $\ll$ zk-STARK, and
Bulletproof proving should be at least five times faster than either alternative. Peer verification
should be the largest phase of Table~\ref{tab:fl-context}. \texttt{DKG3} should be the steepest curve of Figure~\ref{fig:n-vs-time}, and
\texttt{PartialGen} should be flat, as generating a partial decryption does not depend on the
participant count.

Run-to-run variation is small except for the zk-STARK, which uses random sampling. The
Bulletproof and zk-SNARK sizes are deterministic at exactly 924 and 1{,}981 bytes, and on
one machine their measured times vary by a few percent across runs. The mean zk-STARK
size stays within 18{,}500--19{,}100 bytes. Its proving time is the one figure that is not
stable from run to run. The standard deviation across the thirty runs is comparable to
the mean, and the slowest run is around fifty times the fastest, so thirty samples do not
give a stable estimate of the mean. Expect two runs to disagree by roughly 20\% on that cell, and
occasionally by more than half, and expect the zk-STARK round time of Table~\ref{tab:baseline} to inherit
the spread. 

Fixture generation time grows as roughly the cube of the participant count.  Counts of 
5, 10, 20, 50, and 100 are tested by default with a threshold of $1/2$. Run
\texttt{ZKDEAP\_SIZES=5,10,20,50,100,250,500 ./run\_all.sh} for the full range of Figure~\ref{fig:n-vs-time}.

\subsection{Notes}

As Section~\ref{sec:eval-scope} describes, the artifact measures each operation directly. The
round-time figures of Table~\ref{tab:baseline} and the communication totals of Table~\ref{tab:comm} are composed
arithmetically from those measurements. The artifact runs on one core, so every row of Table~\ref{tab:fl-context} is a single-threaded cost.

\end{document}